\documentclass[preprint2]{aastex6}
\usepackage{color}
\usepackage[titletoc]{appendix}
\usepackage[fleqn]{amsmath}
\usepackage{mathtools}
\usepackage{float}
\usepackage{comment}
\usepackage{enumitem}
\usepackage{natbib}
\usepackage{bm}
\usepackage{totcount}
\usepackage{csvsimple}

\newtotcounter{citnum} 
\def\oldbibitem{} \let\oldbibitem=\bibitem
\def\bibitem{\stepcounter{citnum}\oldbibitem}

\shortauthors{Millholland et al.}
\shorttitle{Tidal Inflation of Sub-Saturns}

\begin{document} 

\defcitealias{2019ApJ...886...72M}{M19}

\def\SPSB#1#2{\rlap{\textsuperscript{#1}}\SB{#2}}
\def\SP#1{\textsuperscript{#1}}
\def\SB#1{\textsubscript{#1}}

\title{Tidal Inflation Reconciles Low-Density Sub-Saturns with Core Accretion}
\author{Sarah Millholland$^{1,2}$, Erik Petigura$^3$, Konstantin Batygin$^4$ \\}
\affil{$^1$Department of Astronomy, Yale University, New Haven, CT 06511, USA \\
$^3$Department of Physics \& Astronomy, University of California Los Angeles, Los Angeles, CA 90095, USA \\
$^4$Division of Geological and Planetary Sciences, California Institute of Technology, Pasadena CA 91125, USA}
\altaffiliation{$^2$NSF Graduate Research Fellow}
\email{sarah.millholland@yale.edu}

\begin{abstract}
While the Solar System contains no planets between the sizes of Uranus and Saturn, our current exoplanet census includes several dozen such planets with well-measured masses and radii. These sub-Saturns exhibit a diversity of bulk densities, ranging from $\sim0.1-3 \ \mathrm{g \ cm}^{-3}$. When modeled simply as hydrogen/helium envelopes atop rocky cores, this diversity in densities translates to a diversity in planetary envelope fractions, $f_{\mathrm{env}} = M_{\mathrm{env}}/M_p$ ranging from $\sim10\%$ to $\sim50\%$. Planets with $f_{\mathrm{env}} \approx 50\%$ pose a challenge to traditional models of giant planet formation by core-nucleated accretion, which predict the onset of runaway gas accretion when $ M_{\mathrm{env}} \sim M_{\mathrm{core}}$. Here we show that many of these apparent $f_{\mathrm{env}} \approx 50\%$ planets are less envelope rich than they seem, after accounting for tidal heating. We present a new framework for modeling sub-Saturn interiors that incorporates envelope inflation due to tides, which are driven by the observed non-zero eccentricities, as well as potential obliquities. Consequently, when we apply our models to known sub-Saturns, we infer lower $f_{\mathrm{env}}$ than tides-free estimates. We present a case study of K2-19 b, a moderately eccentric sub-Saturn. Neglecting tides, K2-19 b appears to have $f_{\mathrm{env}} \approx 50\%$, poised precariously near the runaway threshold; by including tides, however, we find $f_{\mathrm{env}} \approx 10\%$, resolving the tension. Through a systematic analysis of $4-8 \ R_{\oplus}$ planets, we find that most (but not all) of the similarly envelope-rich planets have more modest envelopes of $f_{\mathrm{env}} \approx 10\%-20\%$. Thus, many sub-Saturns may be understood as sub-Neptunes that have undergone significant radius inflation, rather than a separate class of objects. Tidally-induced radius inflation likely plays an important role in other size classes of planets including ultra-low-density Jupiter-size planets like WASP-107 b.
\end{abstract}

\section{Introduction}
\label{intro}
Extrasolar systems harbor many planets with no Solar System analogue. The Solar System is devoid of planets with radii between those of Earth and Neptune ($3.9 \ R_{\oplus}$), yet this is the most prevalent extrasolar class size known \citep{2012ApJS..201...15H, 2013ApJ...766...81F, 2013ApJ...767...95D, 2013PNAS..11019273P}. 
It also lacks planets with sizes between Uranus ($4.0 \ R_{\oplus}$) and Saturn ($9.4 \ R_{\oplus}$). These sub-Saturns, which we define as in previous works to be planets in the range $4.0-8.0 \ R_{\oplus}$ \citep{2017AJ....153..142P}, are less common than sub-Neptunes but are still found within $P<300$ days around $\sim7$\% of Sun-like stars \citep{2018AJ....155...89P, 2019AJ....158..109H}.
The lack of local analogues for these short-period exoplanets results in many mysteries about their interior structures, formation, and evolution. One such example is found in the population of sub-Saturns. 

The low densities of sub-Saturns imply that they must have substantial atmospheric envelopes \citep{2010ApJ...712..974R, 2015ApJ...801...41R, 2015ApJ...806..183W}. Their composition is likely a heavy element core of rocky material, surrounded by a gaseous envelope dominated by hydrogen and helium  \citep{2014ApJS..210...20M, 2014ApJ...783L...6W, 2015ApJ...801...41R}. Sub-Saturns as a population exhibit significant scatter in their mass-radius relationship \citep{2017AJ....153..142P}. 

The observed masses and radii may be converted into interior structure constraints by employing core-envelope models. \cite{2014ApJ...792....1L} constructed model planets consisting of Earth-composition cores surrounded by low-density envelopes of H/He. Considering a grid of $M_{\mathrm{core}}$, $M_{\mathrm{env}}$, age, and incident stellar flux, they computed the planetary radius evolution in response to various sources of envelope heating and cooling.\footnote{These sources include cooling of atmosphere through radiation, cooling of the core (which delays the envelope cooling/contraction), and heating from radioactive decay. Within the framework of these standard models, the planet radius is an effective proxy for the envelope mass fraction \citep[e.g.][]{2014ApJ...792....1L, 2016ApJ...831..180C}.} Critically, given these assumptions, one may invert these models and translate the masses and radii of observed planets into constraints on their core and envelope masses. 

Using this approach, there has emerged a subset of low-density sub-Saturns with atmospheric envelopes comprising $\sim 50\%$ of their total mass. (That is, their envelope mass fractions are $f_{\mathrm{env}} \equiv M_{\mathrm{env}}/M_p \approx 50\%$.) Examples of such planets, which are perplexing for reasons we will describe below, include the recently discovered K2-24 c \citep{2018AJ....156...89P} and K2-19 b \citep{2020AJ....159....2P}. Both planets are found near or in mean-motion resonances (MMRs) with neighboring companions and have non-zero eccentricities, $e\sim0.08$ for K2-24 c and $e\sim0.2$ for K2-19 b. Using the \cite{2014ApJ...792....1L} models, \cite{2018AJ....156...89P} estimated that K2-24 c required an envelope mass fraction equal to $f_{\mathrm{env}} = 52^{+5}_{-3}\%$, while K2-19 b was inferred to have $f_{\mathrm{env}} = 44\pm3\%$ \citep{2020AJ....159....2P}. 

These near-$50\%$ envelope mass fraction planets are puzzling because they should be rare according to the theory of core accretion, which represents the dominant paradigm of planet formation for gas-rich planets \citep[e.g.][]{1980PThPh..64..544M, 1982P&SS...30..755S, 1996Icar..124...62P}. Core accretion theory holds that solid cores that form before the dissipation of the protoplanetary disk will accrete gas at a rate set by their internal cooling \citep[e.g.][]{2015ApJ...811...41L, 2016ApJ...825...29G}. When the mass in the envelope reaches a threshold approximately equal to the core mass, the envelope will hydrodynamically collapse under its own self-gravity and undergo a phase of runaway gas accretion at a rate limited by the accretionary mass flux \citep{2005Icar..179..415H, 2016ApJ...829..114B}.

While, for an individual planet, one cannot rule out that the protoplanetary disk disappeared at the same instant that the growing envelope mass reached the core mass, such fine-tuning suggests this should be rare. Thus, the discovery of planets with these exact properties is in tension with this picture. Moreover, the problem is not just that there are specific cases that are in conflict; for $P<100$ days, sub-Saturns appear roughly as common as gas giants. That is, the occurrence rate is flat beyond $\sim 4 \ R_{\oplus}$ \citep{2018AJ....155...89P}. 



In this paper, we propose that most sub-Saturns are not as gas-rich as their masses and radii would suggest, but rather, their envelopes have been inflated through tidal heating. Sub-Saturns are often found on short-period orbits with moderate eccentricities \citep{2017AJ....153..142P}. Short-period planets experience substantially stronger tidal interactions with their host stars than any of the Solar System planets due to a steep inverse distance dependence in the tidal forcing. Tidal heating inevitably arises both from eccentricity tides, which result from the non-uniform tidal forces along the eccentric orbits, and from obliquity tides, which are produced when the planet has a non-zero axial tilt (``obliquity'') of its spin axis relative to its orbital axis. For instance, at 0.1 AU, these tidal interactions can result in a tidal luminosity up to $L_{\mathrm{tide}} \sim 10^{29} \ \mathrm{erg \ s^{-1}}$, or roughly $L_{\mathrm{tide}}/L_{\mathrm{irr}} \sim 0.01$, where $L_{\mathrm{irr}}$ is the incident stellar power (see Figure \ref{Edot_tide_heatmap} below).

Radius inflation from tidal heating is not a new consideration. It was one of the proposed sources of the extra heating required to explain the distended radii of hot Jupiters, whose sizes are infamously at odds with standard thermal evolution models \citep[e.g.][]{2001ApJ...548..466B, 2004ApJ...610..477O, 2008ApJ...681.1631J, 2009ApJ...700.1921I, 2009ApJ...702.1413M}. Tidal heating has also been suggested to be important during the formation of super-Earths/sub-Neptunes by inhibiting their cooling and gas accretion \citep{2017MNRAS.464.3937G}. More recently, \cite{2019ApJ...886...72M}, hereafter \citetalias{2019ApJ...886...72M}, examined the impacts of tidal heating on the radii of sub-Neptunes. They demonstrated that tides can inflate sub-Neptune radii by up to a factor of two when $L_{\mathrm{tide}}/L_{\mathrm{irr}} \gtrsim 10^{-5}$. In particular, \citetalias{2019ApJ...886...72M} proposed that obliquity tides might explain signatures of radius enhancement of planets wide of first-order MMRs.

Planets with active tidal heating are larger at a fixed $f_{\mathrm{env}}$. Consequently, the inclusion of tidal heating in a structural model will yield smaller $f_{\mathrm{env}}$ estimates for observed planets. For instance, \citetalias{2019ApJ...886...72M} presented case studies of some of the ultra-low density ($\rho \lesssim 0.1 \ \mathrm{g \ cm^{-3}}$) ``super-puff'' planets \citep[e.g.][]{2014ApJ...785...15J} and showed that these planets could have $f_{\mathrm{env}}\sim5\%$ (as opposed to $f_{\mathrm{env}}\gtrsim 30\%$) if tidal heating is active. For short-period orbits, tidal inflation and the associated modifications to the $f_{\mathrm{env}}$ estimates are inevitable when the eccentricities and/or obliquities are non-zero.

In this work, we extend the \citetalias{2019ApJ...886...72M} analysis to sub-Saturns by considering planets with larger masses and envelope mass fractions (which were capped in \citetalias{2019ApJ...886...72M} at $20 \ M_{\oplus}$ and 30\%, respectively). We address the question of whether tidal inflation solves the mystery of the apparently anomalous envelope mass fractions, and we study how much tidal inflation impacts the population of sub-Saturns as a whole. We begin with a description of our tidal model, planetary thermal evolution model, and procedure for parameter estimation of observed planets (Section \ref{methods}). We then present a case study of the K2-19 system (Section \ref{K2-19 case study}), which was the original motivation of this work. We expand the analysis to the broader sub-Saturn population in Section \ref{population analysis} and demonstrate that planets across the population are significantly impacted by tidal inflation. In Section \ref{discussion} we discuss the relationship between sub-Saturns and planets in other class sizes, and we consider implications and predictions of our theory.

\section{Methods}
\label{methods}

To study the impacts of tidal heating on the structures of short-period planets, we employ a thermal evolution model that captures the cooling and contraction of the planetary envelope, while also including heating from tidal dissipation. We have done so by building onto a publicly available sub-Neptune evolutionary model developed by \cite{2016ApJ...831..180C}. Here we enumerate our assumptions about tidal dissipation (Section \ref{tidal model}), our thermal evolution model of planetary structure (Section \ref{MESA models}), and our model-fitting and error estimation procedure (Section \ref{MCMC parameter estimation}). Most of these methods were employed in the \citetalias{2019ApJ...886...72M} analysis, and further details may be found therein. Elements of our analysis are available at \href{https://github.com/smillholland/Sub-Saturns/}{https://github.com/smillholland/Sub-Saturns/}; readers can use this code to derive their own planet parameter estimations.

\subsection{Tidal model}
\label{tidal model}

Tidal dissipation involves the conversion of orbital energy into thermal energy in an orbiting body due to tidal deformations raised by the primary body. Tides are complex in general, and the details of where, how, and how much energy is dissipated are non-trivial. Here, we choose to model the tidal dissipation using the viscous approach to traditional equilibrium tide theory \citep[e.g.][]{1880RSPT..171..713D, 1966Icar....5..375G, 1979M&P....20..301M, 1981A&A....99..126H}. The fundamental assumption is that the planet's tidal response to the star is an equilibrium deformation, or tidal bulge, and this bulge lags the star's position with a constant time offset \citep{2008Icar..193..637W, 2008CeMDA.101..171F, 2010A&A...516A..64L}. The physics of the tidal distortion are effectively encapsulated in the parameter $Q'$, the ``reduced tidal quality factor''. While this parameter is highly uncertain for an individual planet, we consider a broad range of plausible values throughout this analysis. 

We note that there are many alternative tidal models that employ different relationships between the tidal forcing frequency and the phase lag angle  \citep[e.g.][]{2009CeMDA.104..257E, 2013CeMDA.116..109F,  2014MNRAS.438.1526S, 2014A&A...571A..50C, 2016CeMDA.126...31B}. Here we are aiming to infer the first-order physical response of tidal heating in sub-Saturn planets, whose specific compositions and rheologies are uncertain. Accordingly, we believe the simplest approach offered by the equilibrium tidal model is an appropriate starting place. 

\begin{figure}
\epsscale{1.2}
\plotone{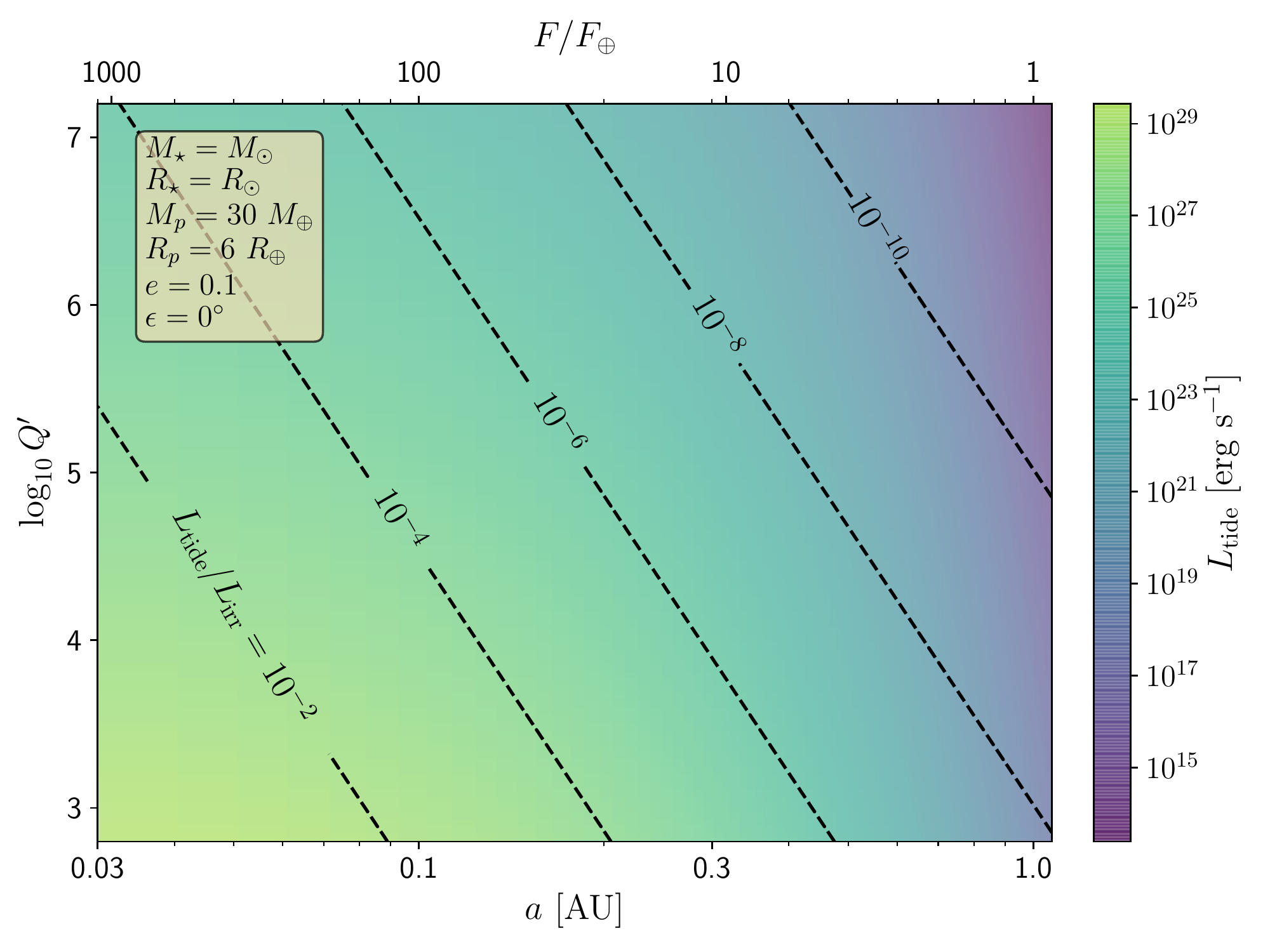} 
\caption{Magnitude of the tidal luminosity, $L_{\mathrm{tide}}$ (equation \ref{full dissipation rate}), as a function of $\log_{10} Q'$ and $a$ (bottom x-axis) or $F/F_{\oplus}$ (top x-axis). We assume fiducial sub-Saturn system parameters, which are shown in the top left. The contours are lines of constant $L_{\mathrm{tide}}/L_{\mathrm{irr}}$, where $L_{\mathrm{irr}}$ is the incident stellar power.}
\label{Edot_tide_heatmap}
\end{figure}

In this framework, the tidal luminosity --- or the rate at which orbital energy is converted into heat energy --- is given by the following expression from \cite{2010A&A...516A..64L}:
\begin{equation}
L_{\mathrm{tide}}(e,\epsilon) = 2 K\left[N_a(e) - \frac{N^2(e)}{\Omega(e)}\frac{2\cos^2\epsilon}{1+\cos^2\epsilon}\right]. 
\label{full dissipation rate}
\end{equation}
Here, $e$ is the orbital eccentricity and $\epsilon$ is the obliquity, or the angle between the planet's spin axis and its orbital axis.\footnote{Note that this angle is distinct from the ``stellar obliquity'' most often referenced in the exoplanetary literature, which is the angle between the stellar spin axis and the planet's orbital axis.} $N_a(e)$, $N(e)$, $\Omega(e)$ are functions of eccentricity given by
\begin{align}
N_a(e) &= \frac{1 + \frac{31}{2}e^2 + \frac{255}{8}e^4 + \frac{185}{16}e^6 + \frac{25}{64}e^8}{(1-e^2)^{\frac{15}{2}}} \label{N_a(e)} \\
N(e) &= \frac{1 + \frac{15}{2}e^2 + \frac{45}{8}e^4 + \frac{5}{16}e^6}{(1-e^2)^6} \label{N(e)} \\
\Omega(e) &= \frac{1 + 3e^2 + \frac{3}{8}e^4}{(1-e^2)^{\frac{9}{2}}}.
\label{Omega(e)}
\end{align}
The quantity $K$ is the characteristic luminosity scale,
\begin{equation}
K = \frac{3n}{2}\frac{k_2}{Q}\left(\frac{G {M_{\star}}^2}{R_p}\right)\left(\frac{R_p}{a}\right)^6, 
\label{tidalK}
\end{equation}
where $n = 2\pi/P$ is the mean motion, $a$ is the semi-major axis, $M_{\star}$ is the stellar mass, and $R_p$ is the planet radius. The two parameters $k_2$ and $Q$ relate to the planet's composition and interior structure. $k_2$ is the dimensionless Love number, and it is connected to the planet's deformation response to tidal disturbance, as well as the central concentration of the planet's density profile.\footnote{For reference, the estimates for Saturn, Uranus, and Neptune are $k_2 = 0.39$, $0.104$, and $0.127$, respectively \citep{1977Icar...32..443G, 2016CeMDA.126..145L}.} $Q = (n\Delta{t})^{-1}$ is the annual tidal quality factor (where $\Delta t$ is the constant tidal time lag), and it parameterizes the efficiency of tidal damping. These parameters are frequently combined into the ``reduced tidal quality factor'', $Q' = 3Q/2k_2$.  

Finally, we have also assumed in equation \ref{full dissipation rate} that the planet's spin rotation frequency, $\omega = 2\pi/P_{\mathrm{rot}}$, has reached an equilibrium (at which $d\omega/dt = 0$). This rate is given by \citep{2010A&A...516A..64L}
\begin{equation}
\omega_{\mathrm{eq}}=n\frac{N(e)}{\Omega(e)}\frac{2\cos\epsilon}{1+\cos^2\epsilon}. 
\label{omega_eq}
\end{equation}
The equilibrium rate reduces to synchronous rotation when $e = 0$ and $\epsilon = 0^{\circ}$.

In order to illustrate typical values of $L_{\mathrm{tide}}$ for sub-Saturns, we plot equation \ref{full dissipation rate} as a heatmap in Figure \ref{Edot_tide_heatmap}. $L_{\mathrm{tide}}$ can reach up to $\sim10^{29} \ \mathrm{erg \ s^{-1}}$, or roughly 1\% of the incident stellar power, $L_{\mathrm{irr}}$. 

\subsection{Planetary thermal evolution modeling}
\label{MESA models}

We now consider how this tidal luminosity affects the structural properties of sub-Saturns. To do so, we employ the model developed by \citetalias{2019ApJ...886...72M} for thermally evolving planetary envelopes while including tidal heating. Details of the model may be found in Section 3 of \citetalias{2019ApJ...886...72M}. In short, we assume a spherically symmetric, two-layer planet model consisting of a heavy element core and an H/He envelope with solar metallicity and helium fraction. The atmospheric envelope is evolved using the Modules for Experiments in Stellar Astrophysics \citep[MESA;][and references therein]{2011ApJS..192....3P, 2013ApJS..208....4P, 2015ApJS..220...15P, 2018ApJS..234...34P} 1D stellar evolution code, including a range of modfications developed by \cite{2016ApJ...831..180C} that make the model specific to planets as opposed to stars. 

The tidal heating is accounted for as an extra source of core luminosity, which amounts to depositing the tidal energy at the base of the atmospheric envelope. This approach was justified in \citetalias{2019ApJ...886...72M} on the basis that solid cores likely have significantly lower tidal quality factors than the planetary envelopes they host \citep{2014MNRAS.438.1526S, 2015MNRAS.450.3952S}, although \citetalias{2019ApJ...886...72M} also showed that the results are nearly identical as long as the heat is deposited at or below the atmosphere's radiative-convective boundary. Qualitatively, this picture is similar to the hot Jupiter radius inflation problem \citep[e.g.][]{2003ApJ...592..555B, 2011ApJ...738....1B, 2013ApJ...772...76S, 2017ApJ...844...94K}. We note, however, that we do not account for Ohmic dissipation or any other additional heating sources that are relevant for planets with $T_{\mathrm{eq}} \gtrsim 1000$ K. This will lead to underestimated radius inflation for the few strongly irradiated planets that we will study in Section \ref{population analysis}; neglecting this effect does not undermine our primary conclusions, but rather strengthens them.

In order to determine the magnitude of tidally induced radius inflation over a wide range of parameter space, \citetalias{2019ApJ...886...72M} developed a suite of $\sim 5,000$ planetary models varying in four principal parameters: the planet mass, $M_p$; the fraction of mass in the H/He envelope, $f_{\mathrm{env}}$; the strength of the incident stellar radiation flux with respect to Earth's, $F/F_{\oplus}$; and the strength of the tidal dissipation, which was parameterized as 
\begin{equation}
\Gamma \equiv \log_{10}\left[\frac{Q'(1+\cos^2\epsilon)}{\sin^2\epsilon}\right] 
\label{Gamma definition}
\end{equation}
via equation \ref{full dissipation rate} with $e=0$. We note that $\log_{10}L_{\mathrm{tide}}$ or $\log_{10}(L_{\mathrm{tide}}/L_{\mathrm{irr}})$ would be a more natural parameter choice than $\Gamma$. Using $\Gamma$ maintains consistency with \citetalias{2019ApJ...886...72M}, where the goal was to isolate obliquity tides. We will soon show that the treatment is formally equivalent.

The host star was assumed to have solar properties, such that $F/F_{\oplus} = (a/{\mathrm{AU}})^{-2}$. The four principal parameters were randomly selected from log-uniform distributions across a wide range of parameter space. For each set of parameters, two models were generated and evolved for 10 Gyr: one model including tidal heating and one without. The planet radii in each case were directly compared. 

The simulations in \citetalias{2019ApJ...886...72M} were confined to $M_p < 20 \ M_{\oplus}$ and $f_{\mathrm{env}} < 30\%$. Here, we must extend the parameter space. We generated $\sim 10,000$ new models, with the ranges of the four principal parameters listed in Table~\ref{MESA model parameters}. We note that while $\Gamma$ is still parameterized using the $e=0$ case, we will later use a transformation of this parameter to generalize the simulations to $e\ne0$. 

\begin{table}[h]
\centering
\caption{Parameters and their ranges used for the set of planet models.}
\begin{tabular}{c | c}
Parameter & Range \\
\hline
$M_p/M_{\oplus}$ & (1, 70)    \\
$\log_{10}{f_{\mathrm{env}}}$ & (-2.5, -0.3) \\
$\log_{10}F/F_{\oplus}$ & (0, 3) \\
$\Gamma$ & (3, 7) \\
\end{tabular}
\label{MESA model parameters}
\end{table}

\subsection{MCMC parameter estimation}
\label{MCMC parameter estimation}
With this expanded set of models in hand, the goal is to make inferences about the structures of observed planets after accounting for radius inflation due to tidal heating. To do this, we use the Markov Chain Monte Carlo (MCMC) approach described in Section 5.1 of \citetalias{2019ApJ...886...72M}. A demonstration of the procedure is provided at \href{https://github.com/smillholland/Sub-Saturns/}{https://github.com/smillholland/Sub-Saturns/}. 

In brief, we first use the MESA simulation set to construct linear barycentric interpolation functions for $R_p$ at $1, 2, ..., 10$ Gyr in both the tides and tides-free models. The independent variables are the four principal parameters --- $M_p$, $f_{\mathrm{env}}$, $F_p$, and $\Gamma$ --- in the case with tides and just the first three of these in the tides-free case. We then use the 5 Gyr\footnote{The choice of 5 Gyr is arbitrary; any age past 1 Gyr works equally well. The radii asymptote to equilibrium values by $\sim1$ Gyr as the planets radiate away their heat of formation.} interpolation functions as the radius models and employ the affine invariant ensemble sampler \texttt{emcee} \citep{2010CAMCS...5...65G, 2013PASP..125..306F} to estimate the posterior distributions of the parameters consistent with the planets’ observed radii. We use uniform priors and a Gaussian likelihood function, and we collect 10,000 samples across 200 chains, discarding the first 5,000 samples as burn-in. Convergence is assessed by inspecting the trace plots to see that the chains are well-mixed. We also calculate the Gelman-Rubin statistic \citep{1992StaSc...7..457G} for each parameter and ensure that it is stable and close to 1.

Within the MCMC, we fix $F/F_{\oplus}$ to the observed value for each planet and let $M_p$ float within $3\sigma$ uncertainties. As a result, the only essential free parameter in the tides-free model is $f_{\mathrm{env}}$. $\Gamma$ is a second free parameter in the model including tides. We will use the notation $f_{\mathrm{env, 0}}$ and $f_{\mathrm{env, t}}$ to refer to the envelope mass fractions obtained from parameter inferences in the case without tides and with tides, respectively. 

The output products of this MCMC procedure are the posterior distributions. In the case of the fit that includes tides, we must apply one final transformation to make the results generalizable. This is because the MESA simulations, as noted in Section \ref{MESA models}, assumed $e=0$ and solar parameters. To allow the posterior distribution to correspond to arbitrary $e$ and stellar parameters, we must transform $Q'$ in a way that forces $L_{\mathrm{tide}}$ to be invariant. To illustrate this, we let the subscript ``$i$'' correspond to the initial results using $e=0$ and solar parameters. Explicitly, we note that $M_{\star i} = M_{\odot}$ and $a_i/{\mathrm{AU}} = (F/F_{\oplus})^{-1/2}$. The parameters without subscripts will correspond to the actual system, with arbitrary $e$, $a$, $\epsilon$, and stellar parameters. Using equation \ref{full dissipation rate} and \ref{Gamma definition} and setting $L_{\mathrm{tide}}(e,\epsilon) = L_{\mathrm{tide},i}(e_i=0,\epsilon_i)$, we obtain
\begin{equation}
\begin{split}
Q' &= Q_i'\frac{(1+\cos^2\epsilon_i)}{\sin^2\epsilon_i}\left(\frac{M_{\star}}{M_{\star i}}\right)^2\left(\frac{n}{n_i}\right)\left(\frac{a_i}{a}\right)^6 \\
&\times \left[N_a(e) - \frac{N^2(e)}{\Omega(e)}\frac{2\cos^2\epsilon}{1+\cos^2\epsilon}\right].
\end{split}
\end{equation}
This transformation allows us to take the original posterior distribution in $\Gamma_i = \log_{10}[Q_i'(1+\cos^2\epsilon_i)/\sin^2\epsilon_i]$ and, for a given $e$, $\epsilon$, $M_{\star}$, and $a$ (the latter two of which give us $n$), transform it into a posterior distribution in $\log_{10}Q'$. By working in log space, the Jacobian of the transformation is unity.

In Sections \ref{K2-19 case study} and \ref{population analysis}, we will employ this MCMC fitting procedure to estimate the parameters of planets whose structures may have been significantly altered by tidal inflation.

\section{Case Study of the K2-19 System}
\label{K2-19 case study} 

We begin with a case study of the K2-19 system, since the recent analysis of this system by \cite{2020AJ....159....2P} was the motivation for this work, and the planetary masses and radii have been constrained at the $\sim 5\%$ level. The system contains three known transiting planets: two sub-Saturns K2-19 b and c, with orbital periods of 7.9 days and 11.9 days, respectively, as well as K2-19 d, a close-in Earth-size planet at 2.5 days. The outer two planets are interacting in an eccentric ($e\sim0.2$) 3:2 mean-motion resonance.\footnote{While \cite{2020AJ....159....2P} found that the standard resonant angles all circulated rather than librated, \cite{2020arXiv200304931P} find that a set of angles emerging from the Sessin-Henrard transformation \citep{2013A&A...556A..28B} \textit{do} in fact librate.} See Table \ref{K2-19 parameters} for a list of the relevant parameters.

Using interpolation to the \cite{2014ApJ...792....1L} core-envelope planet model grid (which does not include tidal heating), \cite{2020AJ....159....2P} estimated that planets b and c have large envelope mass fractions, $f_{\mathrm{env, b}} = 44\pm3 \%$ and $f_{\mathrm{env, c}} = 14\pm1 \%$. As discussed in the introduction, planet b's near-50\% inferred envelope fraction presents a particular challenge to core accretion theory. Here we investigate revisions to these estimates when we account for tidal heating.

\setlength{\extrarowheight}{0pt}
\begin{table}
\centering
\caption{Parameters of the K2-19 system from \cite{2020AJ....159....2P}.}
\begin{tabular}{l r}
Parameter & Value \\
\hline
\multicolumn{2}{c}{Star} \\
$M_{\star} \ (M_{\odot})$ & $0.88\pm0.03$ \\
$R_{\star} \ (R_{\odot})$ & $0.82\pm0.03$ \\
$T_{\mathrm{eff}}$ (K) & $5322\pm100$ \\
\multicolumn{2}{c}{Planet b} \\
$P_b$ (days) & $7.9222\pm0.0001$ \\
$e_b$ & $0.20\pm0.03$ \\
$i_b$ (deg) & $91.5\pm0.1$ \\
$F_b \ (F_{\oplus})$ & $87.0\pm9.5$ \\
$M_{p,b} \ (M_{\oplus})$ & $32.4\pm1.7$ \\
$R_{p,b} \ (R_{\oplus})$ & $7.0\pm0.2$\\
\multicolumn{2}{c}{Planet c} \\
$P_c$ (days) & $11.8993\pm0.0008$ \\
$e_c$ & $0.21\pm0.03$ \\
$i_c$ (deg) & $91.1\pm0.1$ \\
$F_c \ (F_{\oplus})$ & $50.6\pm5.6$ \\
$M_{p,c} \ (M_{\oplus})$ & $10.8\pm0.6$ \\
$R_{p,c} \ (R_{\oplus})$ & $4.1\pm0.2$\\
\multicolumn{2}{c}{Planet d} \\
$P_d$ (days) & $2.5081\pm0.0002$ \\
$e_d$ & $0$ (fixed) \\
$i_d$ (deg) & $90.8\pm0.7$ \\
$F_d \ (F_{\oplus})$ & $403.2\pm44.5$\\
$M_{p,d} \ (M_{\oplus})$ & $<10$ \\
$R_{p,d} \ (R_{\oplus})$ & $1.11\pm0.05$\\
\end{tabular}
\label{K2-19 parameters}
\end{table}

\subsection{Inferences of $f_{\mathrm{env}}$ when including tides}
\label{K2-19 MCMC}

\begin{figure}
\epsscale{1.}
\plotone{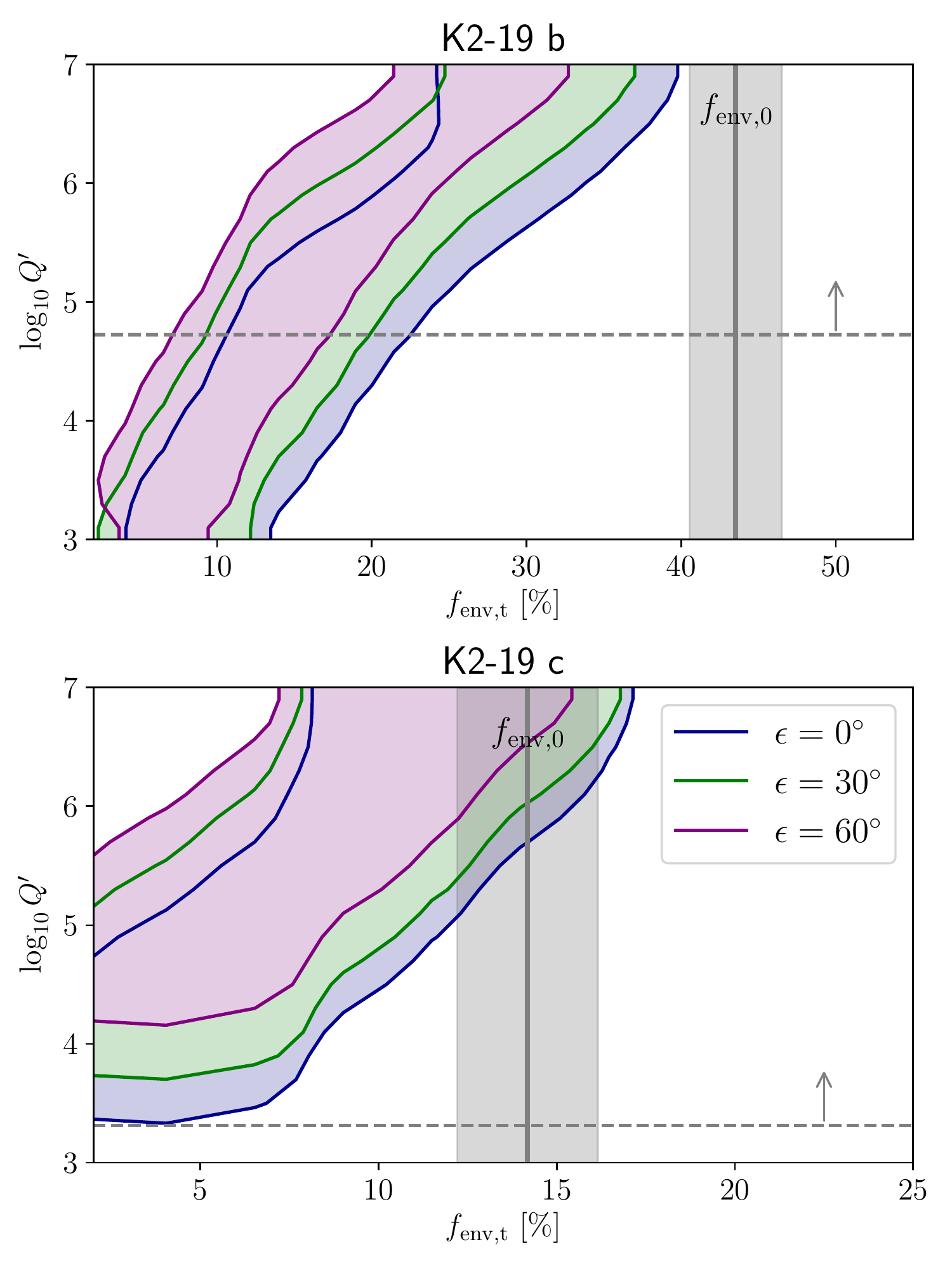}
\caption{Comparison of the envelope mass fraction estimates with tides ($f_{\mathrm{env,t}}$) and without tides ($f_{\mathrm{env,0}}$) of K2-19 b (\textit{top panel}) and K2-19 c (\textit{bottom panel}). The vertical gray bars correspond to the mean and standard deviation of $f_{\mathrm{env,0}}$. These values are $f_{\mathrm{env,0}} = 44.2\pm2.8 \%$ for K2-19 b and $f_{\mathrm{env,0}} = 14.2\pm2.0\%$ for K2-19 c. The colored regions indicate the $2\sigma$ contours of the posterior distributions of $\log_{10}Q'$ and $f_{\mathrm{env,t}}$ after accounting for tides. These assume the eccentricities from Table \ref{K2-19 parameters}, and the separate colors indicate different obliquities ($\epsilon = 0^{\circ}, 30^{\circ}, 60^{\circ}$). The horizontal dashed lines indicate the lower limit $\log_{10}Q'_{\mathrm{low}}$ such that $\tau_e >$ 1 Gyr. } 
\label{K2-19_fits}
\end{figure}

Figure \ref{K2-19_fits} shows the envelope fraction estimates resulting from our MCMC analysis. The top and bottom panels show the results for K2-19 b and K2-19 c, respectively. The mean and 1$\sigma$ range of $f_{\mathrm{env,0}}$ inferred when neglecting tides are shown with the gray line and bar. The tides-free envelope fraction estimates are $f_{\mathrm{env,0}} = 44.2\pm2.8 \%$ for K2-19 b and $f_{\mathrm{env,0}} = 14.2\pm2.0\%$ for K2-19 c, which are both consistent with the estimates from \cite{2020AJ....159....2P}. The agreement is reassuring given that these estimates were obtained with two distinct core-envelope models.

The colored regions in Figure \ref{K2-19_fits} indicate the 2D posterior distributions in $\log_{10}Q'$ and $f_{\mathrm{env,t}}$ from the fit that includes tidal inflation. These distributions assume the measured eccentricities and indicate the results for three values of the obliquity ($\epsilon = 0^{\circ}, 30^{\circ}, 60^{\circ}$). There is a strong covariance between $\log_{10}Q'$ and $f_{\mathrm{env, t}}$; lower values of $Q'$ result in stronger tidal heating, greater amounts of radius inflation, and smaller envelope fraction estimates. 

For planets in this class size, the most reasonable estimate of $Q'$ is in the range $\sim 10^4-10^5$, based on analogy with Saturn \citep{1999ssd..book.....M}, Uranus \citep{1989Icar...78...63T}, and Neptune \citep{2008Icar..193..267Z}, as well as the few sub-Neptune and Neptune-mass exoplanets with $Q'$ constraints. These include the Neptune-mass GJ 436 b \citep[$Q'\sim10^5$,][]{2017AJ....153...86M} and the sub-Neptune-mass GJ 876 d \citep[$Q'\sim10^4-10^5$,][]{2018AJ....155..157P}. However, we can obtain another rough constraint on $Q'$ by considering that the timescale for the orbit to circularize must be greater than the system age, since the planets are observed with non-zero eccentricities today. The tidal circularization timescale is given by \citep{2010A&A...516A..64L}
\begin{equation}
\begin{split}
\tau_e &= \frac{e}{\dot{e}} = \frac{4}{99}\left(\frac{Q'}{n}\right)\left(\frac{M_p}{M_{\star}}\right)\left(\frac{a}{R_p}\right)^5 \\
&\times\left[\Omega_e(e)\cos\epsilon\left(\frac{\omega}{n}\right) - \frac{18}{11}N_e(e)\right]^{-1},
\end{split}
\end{equation}
where we have introduced additional functions of eccentricity (to add to those from equations \ref{N_a(e)}--\ref{Omega(e)}),
\begin{align}
\Omega_e(e) &= \frac{1 + \frac{3}{2}e^2 + \frac{1}{8}e^4}{(1-e^2)^5} \\
N_e(e) &= \frac{1 + \frac{15}{4}e^2 + \frac{15}{8}e^4 + \frac{5}{64}e^6}{(1-e^2)^{\frac{13}{2}}}.
\end{align}

While the stellar age is poorly constrained, we can place an approximate lower limit by assuming $\tau_{\mathrm{age}} \gtrsim 1$ Gyr.\footnote{Consulting gyrochronology relationships, the 20 day rotation period implies an age of $>1$ Gyr (Trevor David, private communication).} Accordingly, we can calculate the lower limit of $Q'$, which we shall denote as $Q'_{\mathrm{low}}$, such that $\tau_e >$ 1 Gyr. In calculating this, we will assume that $\epsilon = 0^{\circ}$ and that $\omega = \omega_{\mathrm{eq}}$. The resulting limits are $Q'_{\mathrm{low}} > 5.3 \times 10^4$ for K2-19 b and $Q'_{\mathrm{low}} > 2.1 \times 10^3$ for K2-19 c. These are indicated with horizontal dashed lines in Figure \ref{K2-19_fits}. We note that the $Q'_{\mathrm{low}}$ estimates are generally conservative, since the system could be younger than 1 Gyr and the planets could have started with significantly larger eccentricities.

We note that there is yet another available constraint on $Q'$ by considering the angular momentum deficit reservoir. That is, the orbital decay induced by tidal dissipation must be balanced by sufficient damping of the eccentricities and mutual orbital inclinations, such that the system's total angular momentum is conserved. This constraint, however, is generally less stringent than the limit imposed by the tidal circularization timescale, since it involves both eccentricity and inclination damping, as opposed to eccentricity damping alone. 

Considering the lower limits on $Q'$, we see that K2-19 b's $f_{\mathrm{env, t}}$ could be as small as $\sim5\%-20\%$, depending on the obliquity. Even with large values of $\log_{10}Q'$, however, we see that the estimate of $f_{\mathrm{env, t}}$ is well below $f_{\mathrm{env, 0}}$, thus resolving the tension produced by the original, tides-free estimate being so near 50\%. For K2-19 c, the impact of tidal heating is less extreme than it is for planet b, but K2-19 c's $f_{\mathrm{env,t}}$ is also well below the tides-free estimate. We also observe that it is possible that the planets have similar intrinsic envelope fractions. For instance, if both planets have $Q' \sim 10^5$, and if $\epsilon_b \sim 60^{\circ}$ and $\epsilon_c \sim 0^{\circ}$, then they would both have $f_{\mathrm{env,t}} \sim 7\%-12\%$. With values this small, planets b and c would both be considered to be sub-Neptunes if they weren't tidally inflated. That is, with $f_{\mathrm{env}}=7\%$ and no tidal heating, the planets would have radii equal to $R_{p,b} \sim 3.9 \ R_{\oplus}$ and $R_{p,c} \sim 3.3 \ R_{\oplus}$. These sizes are at the upper end of the sub-Neptune range ($1.8-4.0 \ R_{\oplus}$, \citealt{2018AJ....155...89P}), suggesting that the K2-19 planets can be interpreted as sub-Neptunes that underwent significant tidal heating.

Lastly, we note that Figure \ref{K2-19_fits} indicates that eccentricity tides alone are sufficient to induce appreciable modifications to the $f_{\mathrm{env}}$ estimates, but obliquity tides may be just as impactful. It is therefore worth investigating the possibility of obliquity enhancement for the K2-19 planets. We address this in Appendix \ref{obliquity excitation for the K2-19 planets}. To summarize, we find that K2-19 b is particularly susceptible to secular spin-orbit resonances that excite the planet's obliquity to large values ($\epsilon \sim 60^{\circ}$). While it is difficult to determine whether this obliquity excitation has in fact transpired, it is clear that obliquity tides may be playing an important role in K2-19 b.

\section{Sub-Saturn Population analysis}
\label{population analysis}

The case study of the K2-19 system has confirmed that tidally induced radius inflation can resolve the near-$50\%$ envelope fraction estimate of K2-19 b, taking it as low as $\sim 5\%-20\%$. It can also reduce K2-19 c's estimate by $\gtrsim 10\%$. This example raises questions about the degree to which other sub-Saturns have been shaped by tidal inflation. In this section, we apply the methodology developed in Section \ref{K2-19 case study} to the sub-Saturn population as a whole.

We begin by defining the planet sample. We use the NASA Exoplanet Archive (NEA, \citealt{2013PASP..125..989A}) to extract all planets with radii in the range $4.0-8.0 \ R_{\oplus}$ that also have measured masses. We remove circumbinary planets and planets with unreliable measurements, such as those with masses based on 10 RVs or less. Where available, we update the planetary and stellar parameters with more precise constraints. We use the  parameter tables from \cite{2018AJ....156..264F}, who combined parallaxes from \textit{Gaia} Data Release 2 (DR2, \citealt{2018A&A...616A...1G}) and spectroscopy from the California-\textit{Kepler} Survey \citep[CKS,][]{2017AJ....154..107P, 2017AJ....154..108J} to update parameters for the \textit{Kepler} planets and planet candidates. 

For the \textit{K2} planets, we use tables from \cite{2020arXiv200111511H}, who similarly derived updated parameters using \textit{Gaia} DR2 and spectroscopy from the Large Sky Area Multi-Object Fibre Spectroscopic Telescope (LAMOST) DR5 \citep{2012RAA....12.1197C}. We only update the stellar and planetary parameters (specifically $R_p$, $R_{\star}$, $a$, $T_{\mathrm{eff}}$, and $F/F_{\oplus}$) when the precision of the $R_p$ measurement is less than the literature value from NEA (which is true for most cases). Finally, for the \textit{Kepler} planets with $M_p$ measured by Transit Timing Variations (TTVs), we use the parameter estimates from the uniform analysis performed by \cite{2017AJ....154....5H} if they are more precise than the NEA masses.

Using the updated masses and radii, we calculate bulk densities and extract the planets with densities measured to 50\% or better. Due to the constraints imposed by the MESA simulations (Section \ref{MESA models}), we can only study planets with $M_p < 70 \ M_{\oplus}$ and $F/F_{\oplus} \in (0,10^3)$. Three planets were removed due to these constraints. The final sample is similar to that assembled by \cite{2017AJ....153..142P}, supplemented by $\sim$15 additional planets that have been discovered since their analysis. Within this list we include TOI-257 b \citep{2020arXiv200107345A} and TOI-421 b \citep{2020arXiv200410095C}, two recently discovered sub-Saturns that had not yet been incorporated into the NEA.

For each sub-Saturn in the sample, we apply the MCMC analysis that was described in Section \ref{MCMC parameter estimation} and utilized for the K2-19 planets in Section \ref{K2-19 MCMC}. We assume fixed eccentricities at the measured values, thus ignoring the effects of any perturbing bodies (observed or unobserved) that might be driving secular eccentricity oscillations and time variable radius inflation. Such oscillations will only affect our inferences at a detailed level. For cases where $e$ was fixed to zero, we assumed a fiducial value equal to $e=0.05$. We obtain envelope fraction estimates of the planets according to the two models with and without tidal inflation. The estimates of $f_{\mathrm{env,0}}$ are calculated using the mean and standard deviation of the posterior distributions from the tides-free fit. The estimates of $f_{\mathrm{env,t}}$ are calculated by first marginalizing the 2D posterior distributions of $f_{\mathrm{env,t}}$ and $\log_{10}Q'$ (e.g. see Figure \ref{K2-19_fits}) by summing them over a range in $\log_{10}Q'$. If $\log_{10}Q'_{\mathrm{low}} < 4$, we use the range $\log_{10}Q' \in [4, 5]$. Otherwise, we use the range $\log_{10}Q' \in [\log_{10}Q'_{\mathrm{low}}$, $\log_{10}Q'_{\mathrm{low}} + 1]$. Finally, we calculate the mean and standard deviation of this marginalized posterior distribution.

\begin{figure}
\epsscale{1.35}
\plotone{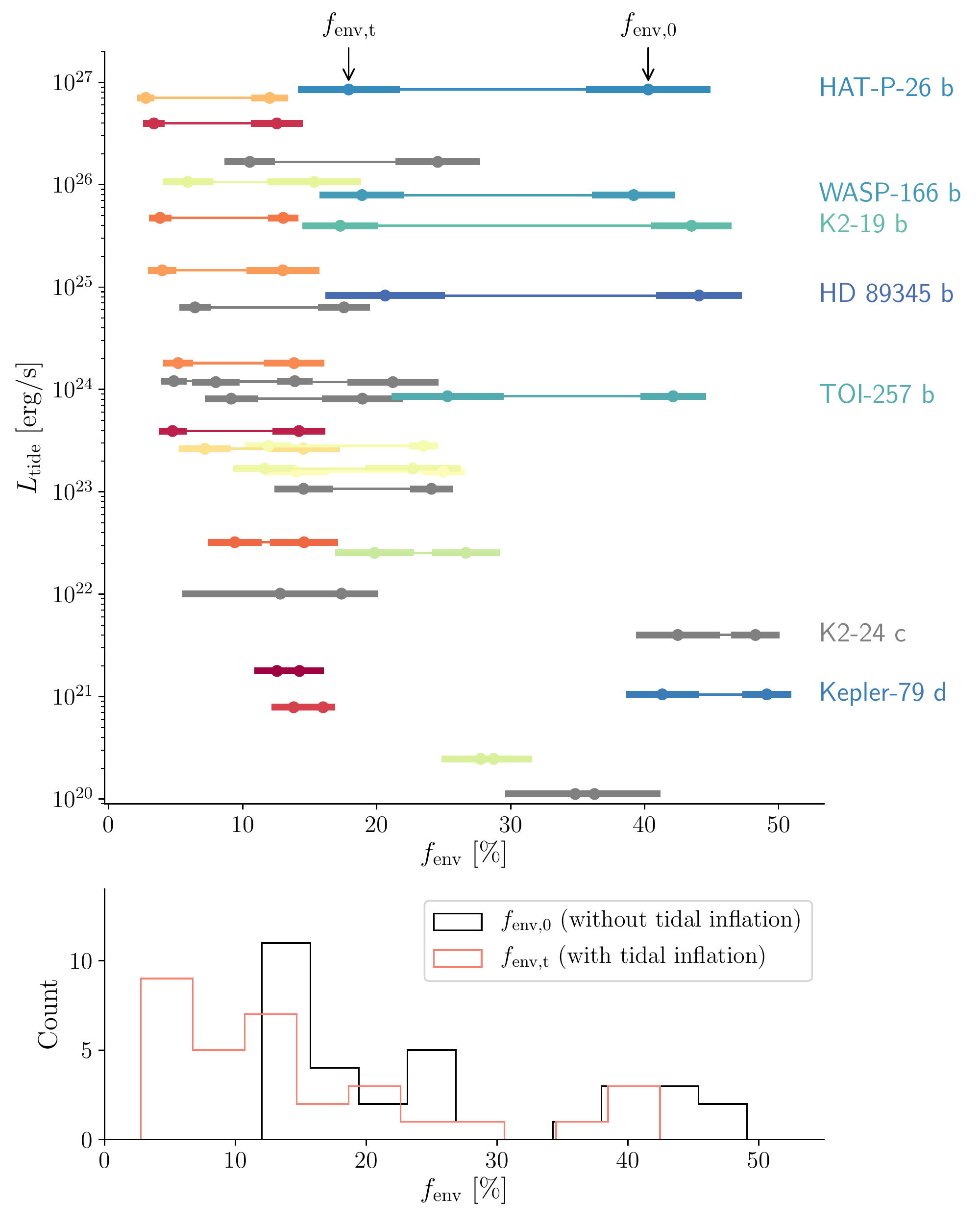}
\caption{Comparison of the envelope mass fraction estimates with tides ($f_{\mathrm{env,t}}$) and without tides ($f_{\mathrm{env,0}}$) across the sub-Saturn population. \textit{Top panel:} On the x-axis we show the two estimates for each planet, where the dot and thick part of the line indicate the posterior mean and standard deviation, respectively. An example is labeled with arrows. The two estimates are connected horizontally in order to indicate the magnitude of the drop in envelope fraction. The color corresponds to $R_p$ (redder = smaller, bluer = larger), and gray lines are the cases with poor eccentricity constraints. The y-axis shows the value of $L_{\mathrm{tide}}$, calculated assuming fiducial values of $Q' = 10^5$ and $\epsilon = 30^{\circ}$. We observe that the change in $f_{\mathrm{env}}$ between the models with and without tides becomes more significant for larger $L_{\mathrm{tide}}$. \textit{Bottom panel:} Histograms of the posterior mean point estimates. Black summarizes the results without tidal inflation, and red summarizes the results with tidal inflation.} 
\label{population_f_env_vs_Ltide_summary}
\end{figure}

The results of this population analysis are summarized in Figure \ref{population_f_env_vs_Ltide_summary} and Table \ref{population_results_table}. For each planet, we show the $f_{\mathrm{env,0}}$ and $f_{\mathrm{env,t}}$ estimates connected by a line. The two estimates are plotted against $L_{\mathrm{tide}}$ in order to examine how the difference in estimates depends on the tidal strength. We observe that most planets have adjustments to their envelope fraction estimates of at least $f_{\mathrm{env,0}} - f_{\mathrm{env,t}} \sim 5\%$; the average value is $\langle f_{\mathrm{env,0}} - f_{\mathrm{env,t}}\rangle = 10\%$. In the bottom panel of Figure \ref{population_f_env_vs_Ltide_summary}, we show histograms of $f_{\mathrm{env,0}}$ and $f_{\mathrm{env,t}}$. Without accounting for tidal inflation, roughly 70\% of planets are inferred to have $f_{\mathrm{env,0}} > 15\%$. In contrast, after accounting for tidal inflation, only 35\% of planets have $f_{\mathrm{env,t}} > 15\%$. 

\begin{figure}
\epsscale{1.2}
\plotone{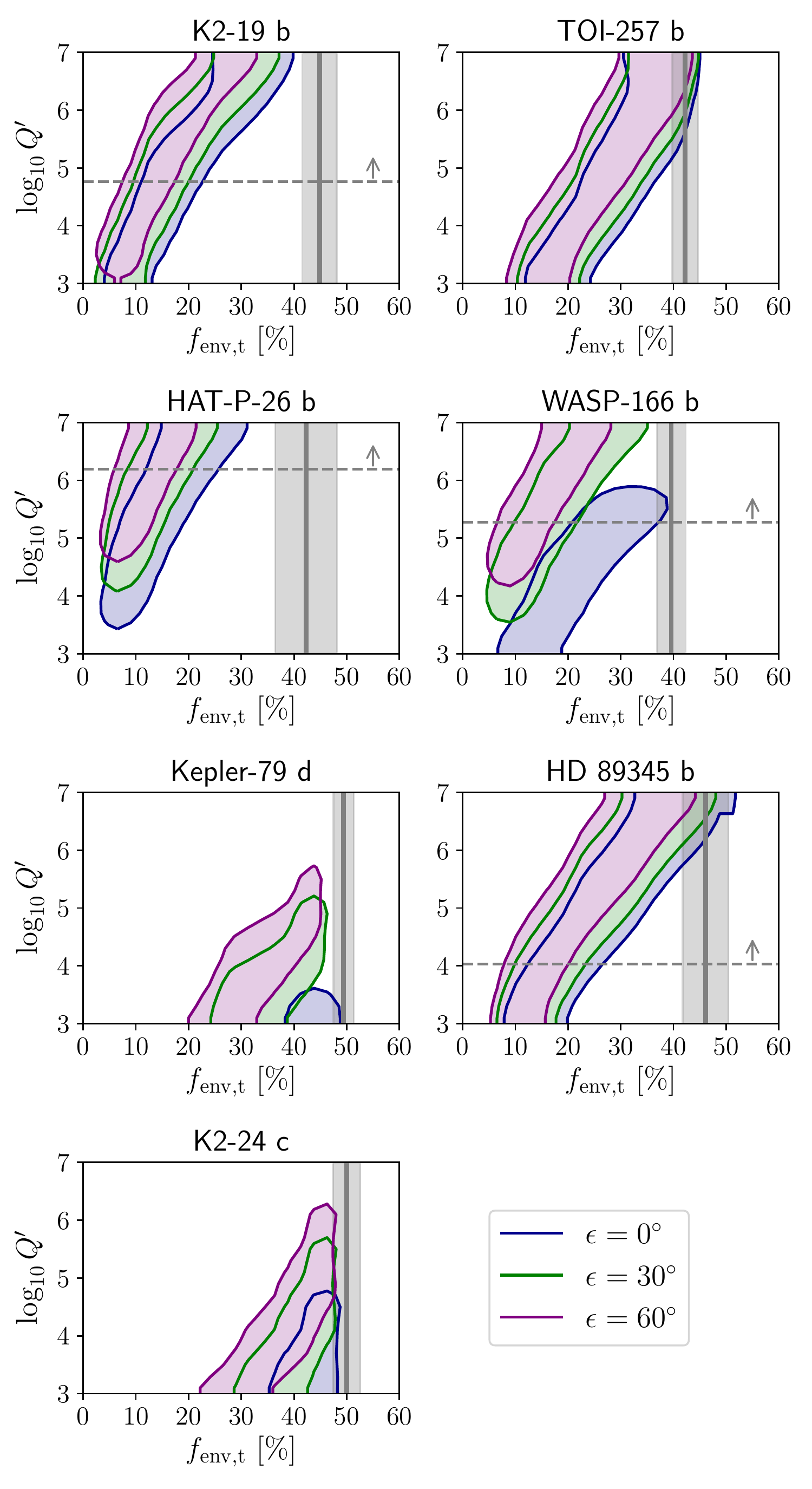}
\caption{Analysis of all planets with $f_{\mathrm{env, 0}} > 40 \%$ before accounting for tidal inflation, including K2-19 b, which was also shown in Figure \ref{K2-19_fits}. As in Figure \ref{K2-19_fits}, we show comparisons of the envelope fraction estimates with tides ($f_{\mathrm{env,t}}$) and without tides ($f_{\mathrm{env,0}}$). The vertical gray bars correspond to the mean and standard deviation of $f_{\mathrm{env,0}}$. The colored regions indicate the $2\sigma$ contours of the posterior distributions of $\log_{10}Q'$ and $f_{\mathrm{env,t}}$ after accounting for tides. These assume the measured eccentricities from Table \ref{population_results_table}, and the separate colors indicate different obliquities. The horizontal dashed lines indicate the lower limit $\log_{10}Q'_{\mathrm{low}}$ such that $\tau_e >$ 1 Gyr. These lines are not pictured for the three cases (TOI-257 b, Kepler-79 d, K2-24 c) where it is below $\log_{10}Q' = 3$.} 
\label{f_env_40_planets}
\end{figure}

The cases of greatest interest are those with $f_{\mathrm{env,0}}~\sim~50\%$. Seven planets have $f_{\mathrm{env,0}}~>~40\%$. These are (from smallest to largest radius) K2-19 b, TOI-257 b, HAT-P-26 b, WASP-166 b, Kepler-79 d, HD 89345 b, and K2-24 c. We plot these cases in Figure \ref{f_env_40_planets}. For convenience of comparison, we include K2-19 b, although it was studied in detail in Section \ref{K2-19 case study}. 

All cases except Kepler-79 d and K2-24 c are unequivocally resolved by including tidal heating, since they have $f_{\mathrm{env,t}}\sim10\%$, even with $\epsilon = 0^{\circ}$. As for Kepler-79 d and K2-24 c, the fit results indicate that both planets would have $f_{\mathrm{env,t}}\sim40\%-50\%$ if $\epsilon = 0^{\circ}$. To get $f_{\mathrm{env,t}} < 30\%$ with tidal inflation would require $\epsilon \gtrsim 30^{\circ}$. Alternatively, there may be other explanations for these planets' anomalously large radii, such as that they are young planets undergoing dusty hydrodynamic outflows \citep{2019ApJ...873L...1W} and/or that they contain high-altitude photochemical hazes that are enhancing their observed radii \citep{2020AJ....159...57L, 2020ApJ...890...93G}.

\setlength{\extrarowheight}{3pt}
\begin{table*}
\caption{Population analysis of sub-Saturns with densities measured to 50\% or better.} 
\label{population_results_table} 
\begin{longtable}{ l  c  c  c  c  c  c  c  c  c}
 \hline
  & Name & $P \ [\mathrm{days}]$ &
  $R_p \ [R_{\oplus}]$ & $M_p \ [M_{\oplus}]$ & $\rho \ [\mathrm{g \ cm}^{-3}]$ & $e$ & $f_{\mathrm{env,0}} \ [\%]$ & $f_{\mathrm{env,t}} \ [\%]$ & $\log_{10}Q'_{\mathrm{low}}$\\
 \hline
 \begin{filecontents*}{population_results_table_050520.csv}
index, plName, Rp, RpLow, RpHigh, Mp, MpLow, MpHigh, rho, rhoLow, rhoHigh, period, periodLow, periodHigh, e, eLow, eHigh, fenv, fenvLow, fenvHigh, fenvt, fenvtLow, fenvtHigh, QprimeLow
1,Kepler-82 b,4.07,0.1,0.24,12.2,0.9,1.0,0.99,0.19,0.11,26.44,0.48,0.48,0.0033,0.0002,0.0019,14.2,1.8,1.8,12.6,1.7,1.7,1.6
2,K2-19 c,4.1,0.2,0.2,10.8,0.6,0.6,0.86,0.14,0.14,11.899,0.001,0.001,0.21,0.03,0.03,14.2,2.0,2.0,4.8,1.0,1.0,3.3
3,GJ 436 b,4.17,0.17,0.17,22.1,2.3,2.3,1.68,0.27,0.27,2.644,0.0,0.0,0.1383,0.0002,0.0002,12.6,2.0,2.0,3.4,0.8,0.8,6.0
4,Kepler-11 e,4.19,0.09,0.07,6.7,1.0,1.2,0.5,0.08,0.1,32.0,0.001,0.001,0.012,0.006,0.006,16.0,0.9,0.9,13.8,1.7,1.7,1.5
5,Kepler-18 c,4.27,0.11,0.11,17.3,1.9,1.9,1.23,0.16,0.16,7.642,0.0,0.0,0.0,0.0,0.0,13.9,1.3,1.3,4.9,1.0,1.0,3.9
6,HD 106315 c,4.35,0.23,0.23,15.2,3.7,3.7,1.02,0.3,0.3,21.057,0.0,0.0,0.22,0.15,0.15,14.6,2.5,2.5,9.4,2.0,2.0,2.2
7,HAT-P-11 b,4.36,0.06,0.06,26.7,2.2,2.2,1.77,0.17,0.17,4.888,0.0,0.0,0.218,0.031,0.034,13.0,1.1,1.1,3.8,0.8,0.8,4.8
8,Kepler-411 c,4.42,0.06,0.06,26.4,5.9,5.9,1.68,0.38,0.38,7.834,0.0,0.0,0.108,0.004,0.003,13.8,2.2,2.2,5.2,1.1,1.1,3.8
9,K2-27 b,4.48,0.23,0.23,30.9,4.6,4.6,1.89,0.41,0.41,6.771,0.0,0.0,0.251,0.088,0.088,13.0,2.7,2.7,4.0,1.1,1.1,4.2
10,Kepler-33 d,4.53,0.14,0.14,3.9,1.8,1.9,0.23,0.11,0.11,21.776,0.0,0.0,0.0,0.0,0.0,17.4,2.6,2.6,12.8,7.3,7.3,2.6
11,Kepler-1656 b,4.57,0.11,0.11,48.6,3.8,4.2,2.8,0.29,0.31,31.579,0.0,0.0,0.836,0.012,0.013,12.0,1.4,1.4,2.8,0.6,0.6,4.7
12,HD 219666 b,4.71,0.17,0.17,16.6,1.3,1.3,0.88,0.12,0.12,6.036,0.001,0.001,0.0,0.0,0.0,17.6,1.9,1.9,6.4,1.2,1.2,4.5
13,Kepler-56 b,4.77,0.21,0.21,22.1,3.6,3.9,1.12,0.24,0.25,10.502,0.001,0.001,0.04,0.01,0.01,14.5,2.7,2.7,7.2,1.9,1.9,3.1
14,K2-32 b,4.96,0.27,0.33,16.5,2.7,2.7,0.75,0.19,0.17,8.992,0.0,0.0,0.0,0.0,0.0,21.2,3.4,3.4,8.0,1.8,1.8,3.9
15,Kepler-18 d,5.11,0.13,0.13,16.4,1.4,1.4,0.68,0.08,0.08,14.859,0.0,0.0,0.0,0.0,0.0,24.1,1.6,1.6,14.5,2.2,2.2,3.0
16,TOI-421 b,5.17,0.13,0.13,16.2,1.1,1.1,0.65,0.07,0.07,16.068,0.0,0.0,0.152,0.045,0.042,25.0,1.6,1.6,14.0,2.4,2.4,3.0
17,K2-98 b,5.2,0.19,0.19,32.2,8.1,8.1,1.26,0.35,0.35,10.136,0.001,0.001,0.0,0.0,0.0,18.9,3.0,3.0,9.1,2.0,2.0,3.4
18,Kepler-25 c,5.22,0.06,0.07,15.2,1.6,1.3,0.59,0.07,0.06,12.721,0.0,0.0,0.0061,0.0041,0.0049,23.5,1.1,1.1,11.9,1.7,1.7,3.4
19,Kepler-223 d,5.24,0.45,0.26,8.0,1.3,1.5,0.31,0.07,0.1,14.789,0.0,0.0,0.037,0.017,0.018,22.7,3.6,3.6,11.6,2.3,2.3,3.5
20,K2-108 b,5.33,0.21,0.21,59.4,4.4,4.4,2.16,0.3,0.3,4.734,0.0,0.0,0.18,0.042,0.042,15.3,3.5,3.5,5.9,1.9,1.9,4.8
21,Kepler-82 c,5.34,0.13,0.32,13.9,1.2,1.3,0.5,0.1,0.06,51.54,0.94,0.94,0.007,0.0018,0.0016,28.7,2.9,2.9,27.8,2.9,2.9,0.9
22,K2-24 b,5.4,0.2,0.2,19.0,2.1,2.2,0.66,0.1,0.11,20.89,0.0,0.0,0.06,0.01,0.01,26.7,2.5,2.5,19.8,2.9,2.9,2.4
23,WASP-156 b,5.72,0.22,0.22,40.7,2.9,3.2,1.2,0.16,0.17,3.836,0.0,0.0,0.0,0.0,0.0,24.6,3.2,3.2,10.5,1.9,1.9,5.5
24,Kepler-87 c,6.14,0.29,0.29,6.4,0.8,0.8,0.15,0.03,0.03,191.232,0.002,0.002,0.039,0.012,0.012,38.8,3.6,3.6,38.6,3.5,3.5,-1.0
25,Kepler-297 c,6.35,0.19,0.19,42.1,9.3,14.1,0.9,0.22,0.31,74.92,0.0,0.0,0.0,0.0,0.0,36.3,4.9,4.9,34.8,5.2,5.2,0.1
26,K2-19 b,7.0,0.2,0.2,32.4,1.7,1.7,0.52,0.05,0.05,7.922,0.0,0.0,0.2,0.03,0.03,44.2,2.8,2.8,17.3,2.8,2.8,4.8
27,TOI-257 b,7.02,0.13,0.15,42.6,7.0,7.3,0.68,0.12,0.12,18.388,0.001,0.001,0.242,0.065,0.04,42.1,2.5,2.5,25.3,4.2,4.2,3.0
28,WASP-166 b,7.06,0.34,0.34,32.1,1.6,1.6,0.5,0.08,0.08,5.444,0.0,0.0,0.03,0.02,0.02,39.2,3.1,3.1,18.9,3.2,3.2,5.3
29,HAT-P-26 b,7.06,0.45,0.45,22.2,6.4,6.4,0.35,0.12,0.12,4.235,0.0,0.0,0.12,0.06,0.06,40.3,4.6,4.6,17.9,3.8,3.8,6.2
30,Kepler-79 d,7.16,0.16,0.13,6.0,1.6,2.1,0.09,0.02,0.03,52.09,0.001,0.001,0.025,0.023,0.059,49.1,1.8,1.8,41.3,2.7,2.7,1.8
31,HD 89345 b,7.4,0.34,0.31,35.0,5.7,5.4,0.48,0.1,0.1,11.814,0.0,0.0,0.22,0.13,0.095,44.1,3.2,3.2,20.6,4.5,4.5,4.0
32,K2-24 c,7.5,0.3,0.3,15.4,1.8,1.9,0.2,0.03,0.03,42.339,0.001,0.001,0.07,0.0,0.0,48.3,1.8,1.8,42.5,3.1,3.1,1.9
\end{filecontents*}
\csvreader[column count = 24, head to column names]{population_results_table_050520.csv}{} 
{\index & \plName & \period &
\Rp\SPSB{$+$\RpHigh}{$-$\RpLow} & \Mp\SPSB{$+$\MpHigh}{$-$\MpLow} & \rho\SPSB{$+$\rhoHigh}{$-$\rhoLow} & \ifcsvstrequal{\e}{0.0}{...}{\ifcsvstrequal{\eLow}{0.0}{$<$\e}{\e\SPSB{$+$\eHigh}{$-$\eLow}}} & \ifcsvstrequal{\fenv}{0.0}{...}{\fenv$\pm$\fenvHigh} &  \ifcsvstrequal{\fenvt}{0.0}{...}{\fenvt$\pm$\fenvtHigh} & \ifcsvstrequal{\QprimeLow}{-1.0}{...}{\QprimeLow} \\}
\end{longtable}
\footnotesize{\textbf{Notes.} $f_{\mathrm{env,0}}$ is the envelope mass fraction estimate when tidal inflation is neglected, and $f_{\mathrm{env,t}}$ is the estimate including tidal inflation. The estimates of $f_{\mathrm{env,t}}$ assumed the measured eccentricities and fiducial $30^{\circ}$ obliquities. $\log_{10}Q'_{\mathrm{low}}$ is a conservative lower limit such that $\tau_e >$ 1 Gyr. (See Section \ref{K2-19 MCMC} for details.) \\
Notes on individual systems: Kepler-82 b --- \cite{2019A&A...628A.108F}, K2-19 c --- \cite{2020AJ....159....2P}, GJ 436 b --- \cite{2014AcA....64..323M}, Kepler-11 e --- \cite{2013ApJ...770..131L}, Kepler-18 c --- \cite{2011ApJS..197....7C}, HD 106315 c --- \cite{2017A&A...608A..25B}, HAT-P-11 b --- \cite{2018AJ....155..255Y}, Kepler-411 c --- \cite{2019A&A...624A..15S}, K2-27 b --- \cite{2017AJ....153..142P}, Kepler-33 d --- \cite{2012ApJ...750..112L}, \cite{2017AJ....154....5H}, Kepler-1656 b --- \cite{2018AJ....156..147B}, HD 219666 b --- \cite{2019A&A...623A.165E}, Kepler-56 b --- \cite{2013Sci...342..331H}, K2-32 b --- \cite{2019A&A...625A..31H}, Kepler-18 d --- \cite{2011ApJS..197....7C}, TOI-421 b --- \cite{2020arXiv200410095C}, K2-98 b --- \cite{2018AJ....156..277L}, \cite{2016AJ....152..193B}, Kepler-25 c --- \cite{2019AJ....157..145M}, Kepler-223 d --- \cite{2016Natur.533..509M}, K2-108 b --- \cite{2017AJ....153..142P}, \cite{2018AJ....156..277L} Kepler-82 c --- \cite{2019A&A...628A.108F}, K2-24 b --- \cite{2018AJ....156...89P}, WASP-156 b --- \cite{2018A&A...610A..63D}, Kepler-87 c --- \cite{2014A&A...561A.103O}, Kepler-297 c --- \cite{2014ApJ...784...45R}, \cite{2014ApJ...787...80H}, K2-19 b --- \cite{2020AJ....159....2P}, TOI-257 b --- \cite{2020arXiv200107345A}, WASP-166 b --- \cite{2019MNRAS.488.3067H}, HAT-P-26 b --- \cite{2011ApJ...728..138H}, \cite{2017AJ....153..136S},  Kepler-79 d --- \cite{2014ApJ...785...15J}, HD 89345 b --- \cite{2018AJ....156..127Y}, K2-24 c --- \cite{2018AJ....156...89P}  }
\end{table*}

\section{Discussion}
\label{discussion}
\subsection{Sub-Saturns as inflated sub-Neptunes}

\begin{figure}
\epsscale{1.2}
\plotone{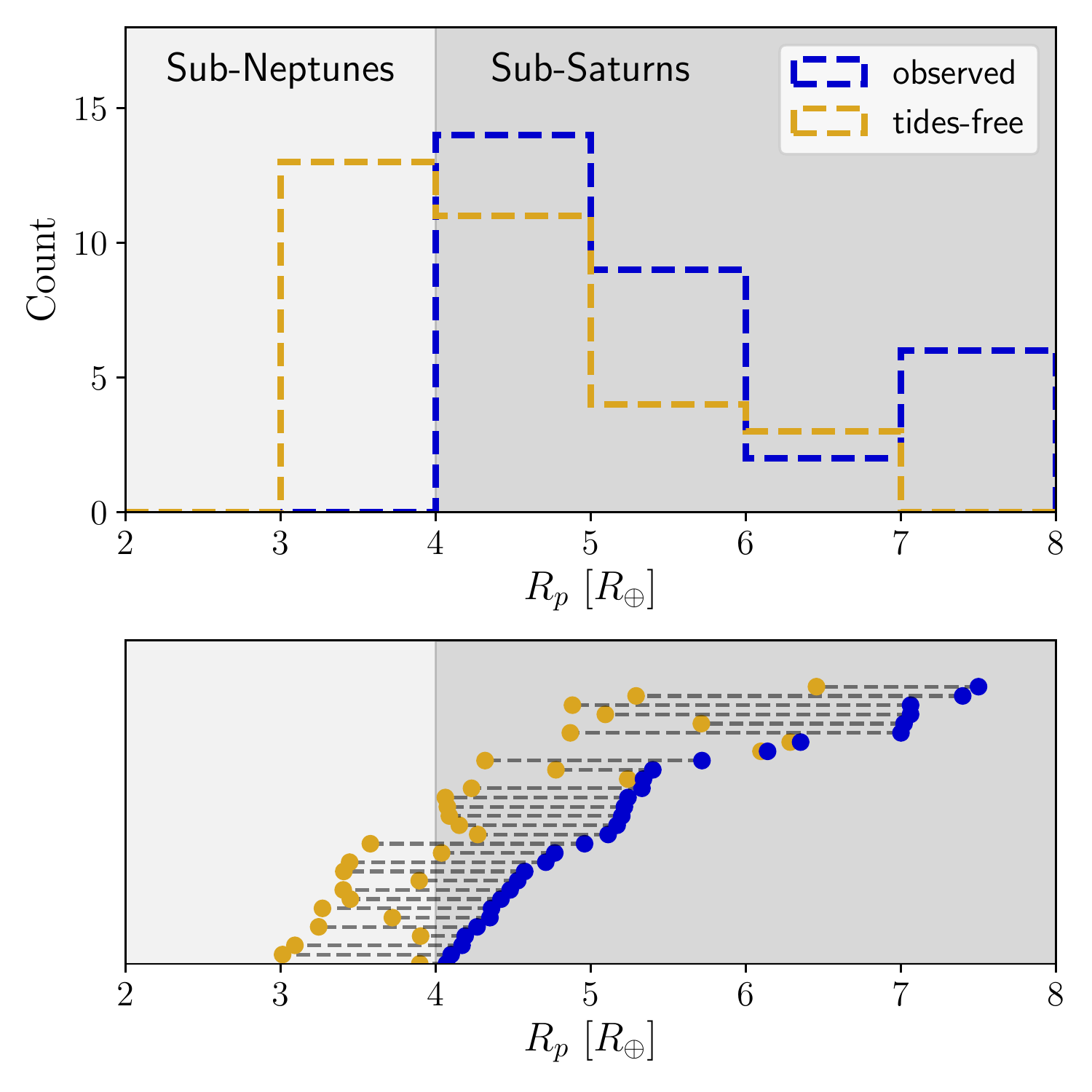}
\caption{Comparison between the observed radii of the planets in the sample (blue) and the radii the planets would have if they were not tidally inflated (yellow), which were calculated assuming $f_{\mathrm{env,t}}$. We illustrate this with histograms (\textit{top panel}) and with point estimates for each planet connected by lines and ordered by the observed radii (\textit{bottom panel}). Nearly half of the planets in the sample would be considered sub-Neptunes ($1.8 - 4.0 \ R_{\oplus}$) if they were not tidally inflated into the sub-Saturn regime ($4.0 - 8.0 \ R_{\oplus}$).} 
\label{Rp_with_and_without_tides}
\end{figure}

We have shown that many sub-Saturns have likely undergone significant radius inflation due to tidal heating. The fact that these planets' radii are influenced by their dynamical state (specifically eccentricity and obliquity) -- and not just their physical properties -- motivates questions surrounding their contextualization within the broader short-period planet population. In other words, many sub-Saturns with $4.0-8.0 \ R_{\oplus}$ may only exist in this radius range because they have been inflated, and their intrinsic envelope fractions may be more consistent with sub-Neptunes, with radii in the range $1.8-4.0 \ R_{\oplus}$. 

To investigate this, we plot in Figure \ref{Rp_with_and_without_tides} the comparison between the observed radii of the planets in the sample and the radii the planets would have if they were not tidally inflated. This estimate is calculated assuming intrinsic envelope fractions equal to $f_{\mathrm{env,t}}$ (from Table \ref{population_results_table}) and using the radius interpolation functions discussed in Section \ref{MCMC parameter estimation}. The observed radii are larger than the tides-free radii by a factor of $1.3$ on average. Nearly half of the planets would be considered sub-Neptunes if they were not experiencing tidal inflation. In this sense, these planets can be interpreted as the tail end of the sub-Neptune group, which have $f_{\mathrm{env}}\lesssim10\%$.

\subsection{Tidal inflation beyond $8.0 \ R_{\oplus}$}
We emphasize that the $4.0 \ R_{\oplus}$ lower range of the sub-Saturn size class is arbitrary, chosen to be a round number near the size of Uranus/Neptune. Similarly, the $8.0 \ R_{\oplus}$ upper bound is an arbitrary classification. For RV surveys seeking to study sub-Jovian mass planets discovered by transit missions \citep[e.g.][]{2017AJ....153..142P}, $8.0 \ R_{\oplus}$ is a convenient threshold. Above this boundary, there is significant ``contamination'' of Jovian-mass planets. However, we should consider whether some planets above the $8.0 \ R_{\oplus}$ threshold may also be experiencing tidal inflation.

\begin{figure}
\epsscale{1.2}
\plotone{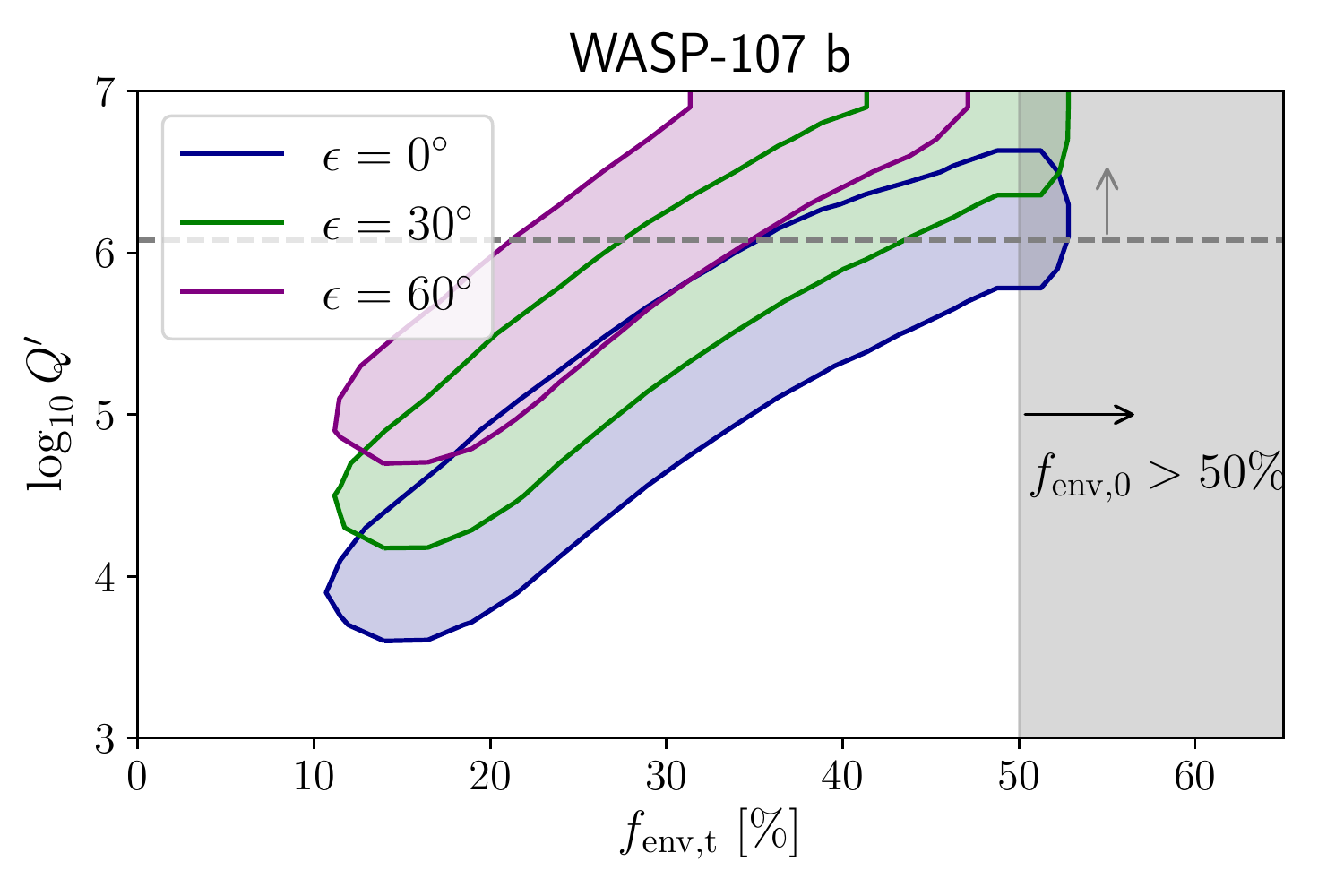}
\caption{Analysis of WASP-107 b, an example of a planet with $R_p$ larger than our $8.0 \ R_{\oplus}$ cutoff, but which nevertheless has likely been substantially affected by tidal inflation due to its high eccentricity ($e=0.13$) and short orbital period ($P=5.7$ days). Similar to Figures \ref{K2-19_fits} and \ref{f_env_40_planets}, the colored regions indicate the $2\sigma$ contours of the posterior distributions of $\log_{10}Q'$ and $f_{\mathrm{env,t}}$ after accounting for tides. These assume the measured eccentricity, and the separate colors indicate different obliquities. The envelope fraction estimate without tides is beyond $f_{\mathrm{env,0}} > 50\%$, which is the upper limit that our model can explore. \cite{2019ESS.....410204P} estimated that it was $f_{\mathrm{env,0}} \sim 80\%$. } 
\label{WASP-107 b}
\end{figure}

One such case is WASP-107 b, a low-density planet with a radius almost as large as Jupiter, $R_p = 10.5 \pm 0.2 \ R_{\oplus}$ \citep{2017A&A...604A.110A}, but with a mass comparable to that of Neptune, $M_p = 30 \pm 2 \ M_{\oplus}$ \citep{2019ESS.....410204P}. It orbits in $P = 5.7$ days with $e = 0.13^{+0.03}_{-0.01}$ \citep{2019ESS.....410204P}. This short-period orbit and non-zero eccentricity creates a substantial tidal luminosity, $L_{\mathrm{tide}} \sim 10^{27} \ \mathrm{erg \ s^{-1}}$ (with $Q' = 10^5$). In Figure \ref{WASP-107 b}, we show the results of our tidal inflation analysis for WASP-107 b, including 2D posteriors in $\log_{10}Q'$ and $f_{\mathrm{env,t}}$. The allowed regime indicated by the posteriors extends as low as $f_{\mathrm{env,t}}\sim10\%$. While the constraint from the circularization timescale suggests that $Q'\gtrsim10^6$, this case study indicates tidal inflation has likely had a significant impact for WASP-107 b and potentially other planets with $R_p > 8.0 \ R_{\oplus}$. We defer a uniform study of this population to future work. 

\subsection{Role of atmospheric mass loss}
There is ample evidence that early atmospheric mass loss, arising from photoevaporation driven by high-energy stellar irradiation \citep[e.g.][]{2017ApJ...847...29O} and/or heating from formation \citep[e.g.][]{2018MNRAS.476..759G}, sculpts the short-period planetary radius distribution, creating a gap between super-Earths and sub-Neptunes \citep{2017AJ....154..109F}. This atmospheric erosion is also expected to impact short-period sub-Saturns, implying that their initial envelope mass fractions would have been even higher than today. 

However, calculations of the instantaneous envelope mass-loss timescale, defined as $\tau_{\mathrm{env}} = M_{\mathrm{env}}/\dot{M_p}$, suggest that sub-Saturns probably did not lose much mass due to their large cores and envelope fractions. For planets with $M_p \gtrsim 10 \ M_{\oplus}$ and $f_{\mathrm{env}} \gtrsim 10\%$, \cite{2016ApJ...831..180C} found that the timescale for photoevaporative mass loss is $\tau_{\mathrm{env}} \approx 10^9-10^{10}$ yr at early times when the stellar EUV is high. This long timescale is in contrast to that for low-mass planets with $M_p \lesssim 10 \ M_{\oplus}$ and $f_{\mathrm{env}} \lesssim 1\%$, which can be as small as $\tau_{\mathrm{env}}\sim10^7$ yr. Consulting our MESA simulations from Section \ref{MESA models}, we confirm that the sub-Saturns did not undergo significant photoevaporative mass loss; when $M_p > 10 \ M_{\oplus}$, $f_{\mathrm{env}}(t=0) > 10\%$, and $F/F_{\oplus} > 50$, the loss in envelope fractions was almost always less than 1\%.

\subsection{Implications for core accretion}

We have presented a resolution for the problematic planets with apparent $f_{\mathrm{env}}\sim50\%$, such that they are no longer situated at the brink of runaway accretion. However, in the process of better understanding their structures, this tidal framework has revealed some additional planets with unusual properties. Four planets in Table \ref{population_results_table} have cores with $M_{\mathrm{core}} > 30 \ M_{\oplus}$ after tidal inflation is accounted for: Kepler-1656 b with 	$M_{\mathrm{core}} = 47.3 \ M_{\oplus}$ and $M_{\mathrm{env}} = 1.3 \ M_{\oplus}$; K2-108 b with $M_{\mathrm{core}} = 55.9 \ M_{\oplus}$	and $M_{\mathrm{env}} = 3.5 \ M_{\oplus}$; 
WASP-156 b with	$M_{\mathrm{core}} = 36.4 \ M_{\oplus}$ and $M_{\mathrm{env}} = 4.3 \ M_{\oplus}$; and TOI-257 b with $M_{\mathrm{core}} = 31.8 \ M_{\oplus}$ and $M_{\mathrm{env}} = 10.8 \ M_{\oplus}$. It is important to note that we would infer these planets to have large cores even when ignoring tides --- $M_{\mathrm{core}} = 42.8 \ M_{\oplus}$, $50.3 \ M_{\oplus}$, $30.7 \ M_{\oplus}$, and $24.7 \ M_{\oplus}$, respectively --- but accounting for tides makes the envelope-to-core contrast even more extreme. These planets probably did not form via core accretion alone; a scattering and high-impact merger event could explain their massive cores, small envelopes, and enhanced eccentricities (particularly Kepler-1656 b with $e=0.836$, as discussed by \citealt{2018AJ....156..147B}). 


Planets with abnormally large cores aside, the majority of sub-Saturns have core masses greater than the $\sim10 \ M_{\oplus}$ critical mass and yet are not gas giants. The problem of how and why they avoided runaway accretion is not unique to sub-Saturns. Many super-Earths and sub-Neptunes also have massive cores that could have reached runaway gas accretion before the dissipation of their disks \citep[e.g.][]{2014ApJ...797...95L}. Various solutions to this have been proposed, including accretion in a high dust-to-gas ratio environment, which slows envelope cooling due to high opacities \citep{2014ApJ...797...95L, 2015ApJ...811...41L}, and hydrodynamical interactions between the envelope of the embedded protoplanet and its disk that slow cooling \citep{2015MNRAS.446.1026O, 2015MNRAS.447.3512O, 2019ApJ...878...36L, 2020arXiv200301644A}. Another potential solution is formation via high-impact protoplanet collisions that remove initial gas atmospheres \citep{2015MNRAS.448.1751I, 2019MNRAS.485.4454B}, leaving the final accretion to occur in a gas-depleted environment. 


\subsection{Influence of atmospheric metallicity}

Our models of the planetary atmospheric envelopes assumed solar metallicities. However, enhanced metallicities have been inferred for many close-in planets ranging from Neptune to Jupiter mass scales \citep[e.g.][]{2017AJ....153...86M, 2019arXiv191108859S}. The increased opacity at high metallicity results in the radiative-convective boundary moving upward in the atmosphere to lower pressures \citep{2019ApJ...884L...6T}, thereby widening the convective zone. This increases the efficiency of tidal heating, such that the radius inflation is more extreme for the same amount of heating. Examining representative case studies with our MESA model, we confirmed that a larger atmospheric metallicity results in both a larger radius and a larger degree of radius inflation for a given $f_{\mathrm{env}}$ and $L_{\mathrm{tide}}$. Accordingly, if most sub-Saturns have enhanced atmospheric metallicities, then they are even more affected by tidal heating than our calculations indicate. 

\subsection{Predictions \& observational signatures}

If many sub-Saturns are in fact less envelope rich than they appear when ignoring tides, we predict that planets with $f_{\mathrm{env}} \sim 50\%$ should not exist (or at least, should be exceedingly rare) on distant orbits where tidal heating is weak, specifically beyond $\sim100$ days. Moreover, unlike the radius occurrence rate distribution, which appears flat beyond $\sim 4 \ R_{\oplus}$, there may be a gap in the distribution of $f_{\mathrm{env}}$ between the primary group of relatively gas poor sub-Neptune/sub-Saturns with $f_{\mathrm{env}} \lesssim 10\%-20\%$ and the gas giant group with $f_{\mathrm{env}} > 100\%$.

In terms of direct observations, the tidal heating in some hot sub-Saturns may exhibit signatures that are potentially observable, particularly with \textit{JWST} \citep{2006SSRv..123..485G}. Tidal heating changes the temperature and chemistry of a planet's deep interior and can result in deviations to its emission spectrum, as demonstrated for GJ 436 b by \cite{2017AJ....153...86M}. Moreover, tidal heating might be observable in full-orbit photometric phase curves through perturbations to the day-night temperature contrasts, since tidal heating is more longitudinally-uniform than stellar insolation. Further modeling work is necessary to investigate this in detail. Finally, signatures of non-zero obliquities, which can induce obliquity tides, can be probed by phase curve modeling \citep{2019AJ....158..108A} and secondary eclipse mapping \citep{2017ApJ...846...69R}. 



\section{Summary}
\label{summary}
Many sub-Saturns have such large radii and low masses that standard core-envelope structural models suggested that they must have massive atmospheric envelopes comprising $\sim50\%$ of their total mass \citep[e.g.][]{2018AJ....156...89P, 2020AJ....159....2P, 2020arXiv200107345A}. This posed a problem for the theory of core accretion, which dictates that planets undergo runaway gas accretion when their envelope mass reaches their core mass. These 50/50 core/envelope planets, however, are generally found on short-period orbits with non-negligible eccentricities. Accordingly, tidal heating from eccentricity tides (as well as potential obliquity tides) is substantial, and this heating generates atmospheric inflation. When this effect is ignored, the atmospheres can appear much more massive than they actually are.  

Beginning with a case study of the K2-19 system,\footnote{Our analysis for K2-19 b, along with a condensed dataset of the MESA simulations and our general procedure for estimating envelope fractions, is provided at \href{https://github.com/smillholland/Sub-Saturns/}{https://github.com/smillholland/Sub-Saturns/}.} we showed that heating from tidal dissipation can resolve the apparent structural problems. K2-19 b's envelope mass fraction estimate is $f_{\mathrm{env,t}}\approx5\%-20\%$ when tides are included in the model, rather than the $f_{\mathrm{env,0}}\approx50\%$ previously thought. Within the broader population of sub-Saturns, estimates of their envelope mass fractions are smaller than previously thought by an average of $\langle f_{\mathrm{env,0}} - f_{\mathrm{env,t}}\rangle = 10\%$. Nearly half of these planets would have radii consistent with the sub-Neptune population if they were not tidally inflated. Thus, a short-period planet's core mass and envelope mass fraction -- rather than its radius, which is dependent on its dynamical state through tidal inflation -- are more fundamental tracers of the planet formation process. Further demographic and modeling studies are required to examine these properties across the planet population.

\section{Acknowledgements}
We thank Chris Spalding for insightful conversations. We also thank Howard Chen \& Leslie Rogers for their publicly available MESA model. S.M. is supported by the NSF Graduate Research Fellowship Program under Grant  DGE-1122492. E.P. acknowledges the generous support of the Alfred P. Sloan Foundation. K.B. is grateful to the David and Lucile Packard Foundation and the Alfred P. Sloan Foundation for their generous support. This research has made use of the NASA Exoplanet Archive, which is operated by the California Institute of Technology, under contract with the National Aeronautics and Space Administration under the Exoplanet Exploration Program.

\appendix

\section{Obliquity excitation from secular spin-orbit resonance}
\label{obliquity excitation for the K2-19 planets}

In Section \ref{K2-19 case study}, we showed that inferences of the envelope mass fractions of the K2-19 planets are significantly impacted after accounting for tidal inflation from eccentricity and/or obliquity tides. The eccentricities are measured to be $e\sim0.2$ for both K2-19 b and K2-19 c, but their obliquities are unknown. The K2-19 planets may have non-zero obliquities if they are trapped in secular spin-orbit resonances, dynamical configurations that involve a commensurability between the frequencies of the planet's spin-axis precession and orbital precession. Encounters with these resonances lead planets to reach high obliquities. The mechanism is responsible for the obliquity excitations of several Solar System planets \citep[e.g.][]{1993Natur.361..608L, 1993Sci...259.1294T, 2004AJ....128.2501W, 2006ApJ...640L..91W}, and it has recently been studied in the context of short-period exoplanetary systems \citep[e.g.][]{2018ApJ...869L..15M, 2019NatAs...3..424M}. In particular, \cite{2019NatAs...3..424M} showed that planets in short-period, compact, multi-planet systems are intrinsically susceptible to these obliquity-exciting resonances due to a proximity between their spin and orbital precession frequencies. Since K2-19 is one such compact multi-planet system, its planets may have enhanced obliquities. To examine this explicitly, here we will calculate the system's precession frequencies and explore their plausible evolutionary history. 

We begin by defining the relevant frequencies of precession of the planets' spin axes and orbits. Planetary spin axis precession arises in response to the torque induced by the host star onto the rotationally-flattened figure of the planet. The precession period is equal to $T_{\alpha}=2\pi/(\alpha\cos\epsilon)$, where $\alpha$ is called the precession constant. In the absence of satellites, $\alpha$ is given by 
\citep{1997A&A...318..975N, 2003Icar..163....1C} 
\begin{equation}
\alpha=\frac{1}{2}\frac{M_{\star}}{M_p}\left(\frac{R_p}{a}\right)^3\frac{k_2}{C}\frac{\omega}{(1-e^2)^{3/2}}.
\label{alpha}
\end{equation}
In addition to the variables defined earlier in Section \ref{tidal model}, $C$ is the planet's moment of inertia normalized by $M_p {R_p}^2$. We may assume that short-period planets have reached their equilibrium spin rotation rate (equation \ref{omega_eq}).

The second relevant precessional motion is the orbit nodal regression, defined with a frequency $g = \dot{\Omega}$, where $\Omega$ is the longitude of ascending node. Nodal recession may be induced by a variety of sources (in general, any deviation from a $1/r$ potential). Here we consider the case that it is driven by secular planet-planet interactions. To be explicit, we delineate the example of an isolated two-planet interaction, but the full Laplace-Lagrange secular solution must be used to capture the dynamics of multi-planet systems. When eccentricities and inclinations are small, the longitudes of ascending node of two planets not near mean-motion resonances will regress uniformly at a frequency \citep{1999ssd..book.....M}
\begin{equation}
g_{\mathrm{LL}} = -\frac{1}{4}b_{3/2}^{(1)}(\alpha_{ij})\alpha_{ij}\left(n_i\frac{M_{pj}}{M_{\star}+M_{pi}}\alpha_{ij}+n_j\frac{M_{pi}}{M_{\star}+M_{pj}}\right).
\label{g_LL}
\end{equation} 
Here the subscript LL stands for Laplace-Lagrange. In addition, $\alpha_{ij}=a_i/a_j$, and we use the subscripts $i$ and $j$ to refer to a pair of planets with planet $i$ the interior of the two. The constant, $b_{3/2}^{(1)}(\alpha_{ij})$, is a Laplace coefficient, defined by 
\begin{equation}
b_{3/2}^{(1)}(\alpha_{ij})=\frac{1}{\pi}\int_{0}^{2\pi}\frac{\cos\psi}{(1-2\alpha_{ij}\cos\psi+{\alpha_{ij}}^2)^{3/2}}d\psi.
\end{equation}
Before proceeding, we briefly note two modifications to $g_{\mathrm{LL}}$. First, if the planets are near MMR, as the K2-19 planets are, the nodal recession frequency is smaller by a factor of $g/g_{\scriptscriptstyle \mathrm{LL}}\sim 0.5$, which depends on the eccentricities and libration amplitude \citep{2019CeMDA.131...38S}. Second, in systems with more than two planets, the orbits will not precess uniformly; instead they will have several eigenfrequency modes within their perturbations. The secular spin-orbit resonance may arise as a commensurability between the spin axis precession frequency and any one of these secular eigenfrequencies \citep[e.g.][]{2004AJ....128.2501W}.

\begin{figure}
\epsscale{0.6}
\plotone{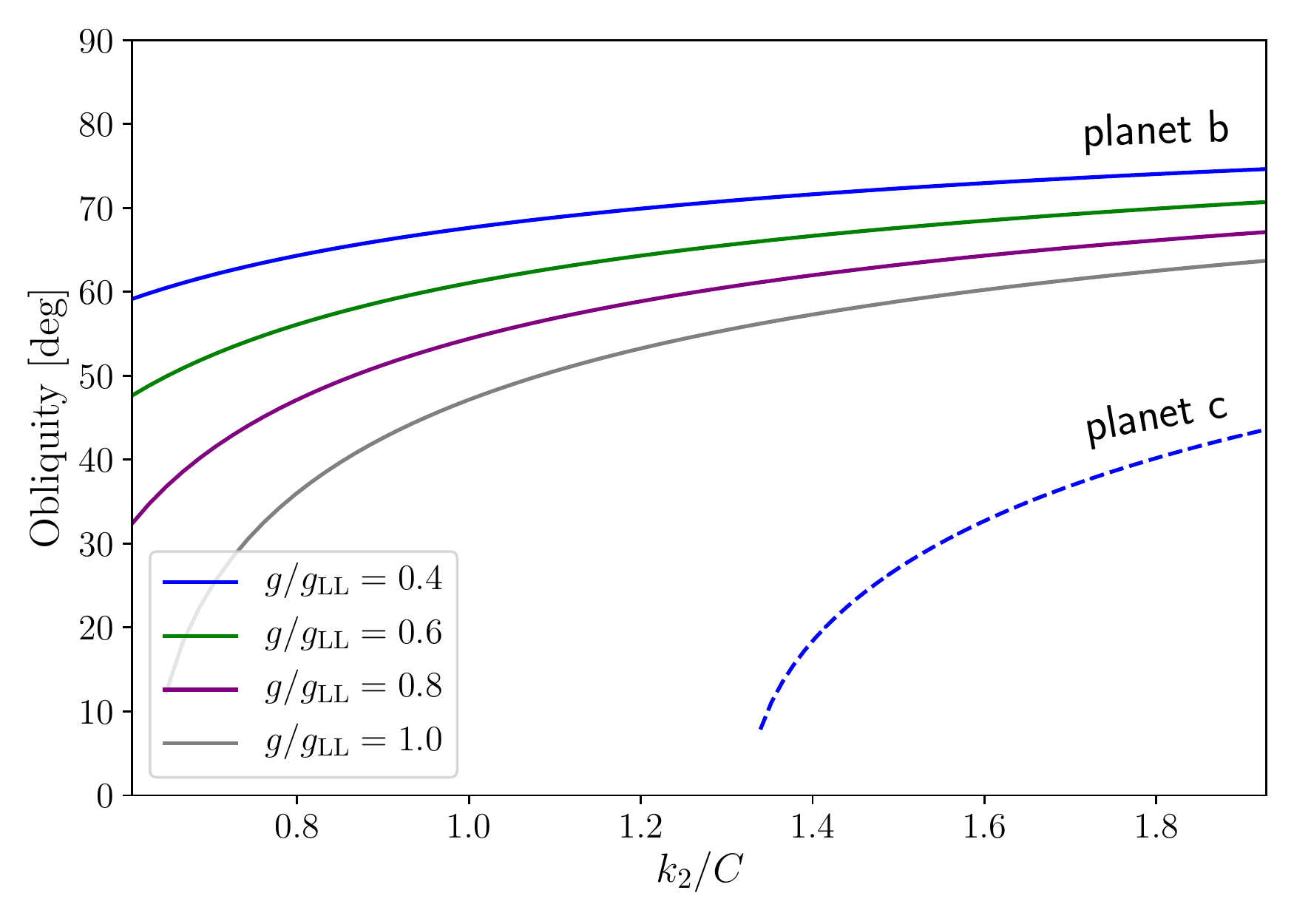}
\caption{The resonant equilibrium obliquity (equation \ref{Cassini state 2 obliquity}) for K2-19 b (solid lines) and K2-19 c (dashed lines) as a function of $k_2/C$ and for different values of $g/g_{\mathrm{LL}}$. A high obliquity resonance for planet b is robust across a range of parameter space; the resonance is possible for planet c but less likely.} 
\label{K2-19_obliquity_range}
\end{figure}

A secular spin-orbit resonance is a particular case of a more general phenomenon called a ``Cassini state'', an equilibrium configuration of the planet's spin vector in the precessing orbital frame. There are up to four Cassini states depending on the value of the ratio $g/{\alpha}$. For a given set of frequencies, $g$ and $\alpha$, and a given orbital inclination with respect to the invariable plane, $I$, the equilibrium solutions of the planet obliquity satisfy 
\begin{equation}
g\sin(\epsilon-I)+\alpha\cos\epsilon\sin\epsilon=0.
\end{equation}
When $I << \epsilon$, this criterion simplifies to 
\begin{equation}
\lvert{g}\rvert\approx\alpha\cos\epsilon.
\label{resonance_condition}
\end{equation}
This simplification is appropriate for the case under consideration here, given the sub-degree mutual inclinations of the planetary orbits (see Table \ref{K2-19 parameters}).

In order to calculate the resonant obliquity -- specifically, the stable equilibrium obliquity in Cassini state 2 -- we can combine equations \ref{omega_eq} and \ref{resonance_condition} and solve for the obliquity. The result is 
\begin{equation}
\label{Cassini state 2 obliquity}
\cos\epsilon = \left[2\frac{N(e)}{\Omega(e)}\frac{\alpha_{\mathrm{syn}}}{{\lvert{g}\rvert}}-1\right]^{-\frac{1}{2}},
\end{equation}
where $\alpha_{\mathrm{syn}} = \alpha(n/\omega)$ is the spin precession constant in the case of synchronous rotation with $\omega = n$. To be clear, this expression yields the obliquity the planet would have if it is resonance, but this resonance locking is not guaranteed.

We use equation \ref{Cassini state 2 obliquity} to determine whether the K2-19 planets could plausibly have excited obliquities due to their participation in secular spin-orbit resonances. All parameters involved in this calculation have been constrained (and are listed in Table \ref{K2-19 parameters}), with the exception of the values of $k_2$ and $C$. We assume fiducial values as follows. Based on estimates \citep{2016CeMDA.126..145L} of the Love numbers for Saturn ($k_2 = 0.39$), Uranus ($k_2 = 0.104$), and Neptune ($k_2 = 0.127$), we consider a broad range of values, $k_2 \in [0.1, 0.5]$. Next, we calculate $C$ by making use of the Radau-Darwin relationship \citep{1984plin.book.....H, 2011A&A...528A..18K}, which relates the non-dimensional moment of inertia to the first-order response to tidal distortion,
\begin{equation}
C_{\mathrm{rd}} = \frac{2}{3}\left[1 - \frac{2}{5}\left(\frac{5}{k_2 + 1} - 1\right)^{\frac{1}{2}}\right].
\end{equation}

Figure \ref{K2-19_obliquity_range} shows curves of the resonant obliquities for planets b and c as a function of $k_2/C$ and $g/g_{\mathrm{LL}}$. This latter quantity determines the degree to which the nodal recession frequency deviates from the Laplace-Lagrange value. We see that planet b could quite easily be in secular spin-orbit resonance; that is, its resonant obliquity is well-defined for a wide range of $k_2/C$ and $g/g_{\mathrm{LL}}$. Planet c could also have a resonantly-enhanced obliquity, but it less likely, as it would require a relatively small value of $g/g_{\mathrm{LL}}$ and $k_2/C \gtrsim 0.4$.

\begin{figure}
\epsscale{0.65}
\plotone{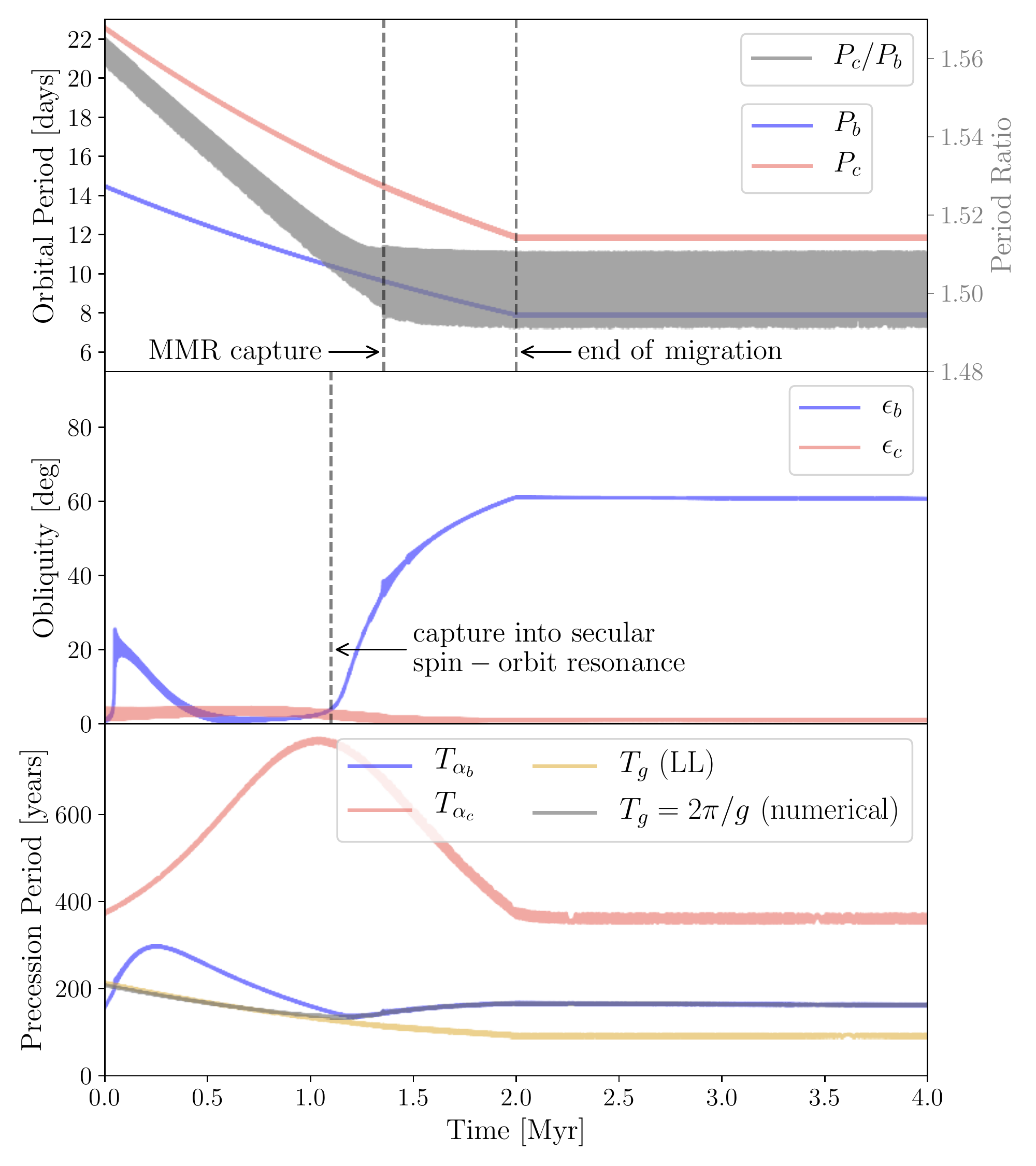}
\caption{Example dynamical evolution of the K2-19 system while planets b and c undergo convergent inward migration for the first 2 Myr. \textit{Top panel:} Evolution of the orbital periods and period ratio; the 3:2 MMR is capture at $\sim 1.3$ Myr, and the migration ends at 2 Myr. \textit{Middle panel:} Evolution of the obliquities of planets b and c. \textit{Bottom panel:} Evolution of the spin and orbital precession periods. The blue and red curves show the precession periods of the spin axes of planets b and c, respectively. The gray curve shows the orbit nodal recession period, which increases away from the Laplace-Lagrange secular solution in yellow as the system enters the MMR.} 
\label{obliquity_excitation_sim}
\end{figure}

We can make this argument more explicit by showing a plausible example of the system’s evolutionary history. To do this, we apply a gravitational $N$-body code that models the planetary tidal, spin, and orbital evolution. This is a direct numerical integration (as opposed to a secular calculation) adopted from the framework of \cite{2002ApJ...573..829M}.  We account for accelerations due to tides raised on the planets from the host star using equilibrium tide theory in the viscous approximation, and we include the accelerations on the planets from the star’s quadrupole gravitational potential. The details of the code are provided in Section 1.2 of the Methods of \cite{2019NatAs...3..424M}.

The basic set-up of our simulation consists of the K2-19 host star and planets b and c, with all bodies endowed with structure. We assign fiducial values, $k_{2,b} = k_{2,c} = 0.3$, $C_b = C_c = 0.3$, and $Q_b = Q_c = 10^4$. We set the initial rotation periods to reflect partial but incomplete spin-down, $P_{\mathrm{rot},b} = 5$ days and $P_{\mathrm{rot},c} = 3$ days. Finally, we initialize the planets on apsidally-aligned orbits with initial eccentricities, $e_1 = e_2 = 0.15$. While the details of the resulting evolution depend somewhat on these initial conditions, the primary features of the response are generalizable.

If the outer planets in the K2-19 system are indeed locked into a mean motion resonance, they must have experienced slow convergent migration. Accordingly, we assume that planets b and c initially start slightly wide of a 3:2 period ratio and migrate inwards convergently due to (unmodeled) interactions with the protostellar disk. We set the migration timescales equal to $\tau_{a_b} = a_b/{\dot{a_b}} = 5$ Myr and $\tau_{a_c} = \tau_{a_b}/1.1$, and the set-up is such that the planets capture the MMR at $\sim 1.3$ Myr and reach their present-day semi-major axes after 2 Myr. The longitudes of perihelion are locked into libration about $\Delta\varpi \approx 0$, in accordance with the observed apsidal alignment of the system.

The resulting dynamical evolution is displayed in Figure \ref{obliquity_excitation_sim}. Planet b first encounters the secular spin-orbit resonance at $\sim 1.1$ Myr when $T_{\alpha}/T_g$ crosses unity from below, producing a resonant kick of the obliquity to $\sim 20^{\circ}$. Shortly thereafter, planet b encounters the resonance again, this time $T_{\alpha}/T_g$ reaching unity from above and resulting in a resonant capture. The planet's obliquity reaches $60^{\circ}$ by the end of the migration at 2 Myr. While planet b's obliquity is readily excited by the resonant interactions, planet c does not encounter the resonance in this example. This is consistent with the constraints indicated in Figure \ref{K2-19_obliquity_range}.

\bibliographystyle{aasjournal}
\bibliography{main}

\end{document}